\documentclass[12pt]{article}
\usepackage{amsmath}
\usepackage{graphicx,psfrag,epsf}
\usepackage{enumerate}
\usepackage{natbib}
\usepackage{url} % not crucial - just used below for the URL 
\usepackage{subfig}
\usepackage{verbatim} 
\usepackage{times}
\usepackage{bm}
\usepackage{caption}
\usepackage[dvips]{epsfig}
\usepackage{amssymb}
\usepackage{graphicx}
\usepackage{indentfirst}
\usepackage[usenames]{color}
\usepackage{float}
\usepackage{lscape}
\usepackage{setspace}
\newcommand{\blind}{0}

\usepackage{caption}                         % <--- added

\newtheorem{theorem}{Theorem}

\newtheorem{lemma}{Lemma}

\begin{document}

\def\spacingset#1{\renewcommand{\baselinestretch}%
{#1}\small\normalsize} \spacingset{1}

%%%%%%%%%%%%%%%%%%%%%%%%%%%%%%%%%%%%%%%%%%%%%%%%%%%%%%%%%%%%%%%%%%%%%%%%%%%%%%

\if0\blind
{
  \title{\bf Optimal-$k$ difference sequence in nonparametric regression}
  \author{Wenlin Dai\\
    Institute of Statistics and Big Data, Renmin University of China,\\ Beijing 100872, China\\
    and \\
    Xingwei Tong\\
    School of Statistics, Beijing Normal University, Beijing 100875, China\\
    and\\
    Tiejun Tong\\
    Department of Mathematics, Hong Kong Baptist University, Hong Kong
    }
  \maketitle 
} \fi

\if1\blind
{
  \bigskip
  \bigskip
  \bigskip
  \begin{center}
    {\LARGE\bf Optimal-$k$ difference sequence in nonparametric regression}
\end{center}
  \medskip
} \fi

\begin{abstract}
\spacingset{1.45}
		Difference-based methods have been attracting increasing attention in nonparametric regression, in particular for estimating the residual variance.
		To implement the estimation, one needs to choose an appropriate difference sequence, mainly between {\em the optimal difference sequence} and {\em the ordinary difference sequence}.
		The difference sequence selection is a fundamental problem in nonparametric regression, 
		and it remains a controversial issue for over three decades. In this paper, we propose to tackle this challenging issue from a very unique perspective, namely by introducing a new difference sequence called {\em the optimal-$k$ difference sequence}.
		The new difference sequence
		not only provides a better balance between the bias-variance trade-off, but also
		dramatically enlarges the existing family of difference sequences that includes the optimal and ordinary difference sequences as two important special cases.
		We further demonstrate, by both theoretical and numerical studies, that the optimal-$k$ difference sequence has been pushing the boundaries of our knowledge in difference-based methods in nonparametric regression, and it always performs the best in practical situations.
		
		\vskip 20pt
		\noindent%
\end{abstract}

\newpage
\spacingset{1.45} % DON'T change the spacing!
\section{Introduction}
	\sloppy{}
	\noindent
	Nonparametric regression models have been widely used in statistics and economics in the past several decades, mainly because of their flexibility in capturing the relationship between the dependent and independent variables.
	%They have been extensively studied during the past several decades \cite{Eubank1999bk}.
	In this paper, we consider the nonparametric regression model
	\begin{equation}
		Y_i=g(x_i)+\varepsilon_i,\quad i=1,\dots, n, \label{Model}
	\end{equation}
	where $\{Y_i\}$ are the observations, $g$ is a mean function, $\{x_i\}$ are the design points, and $\{\varepsilon_i\}$ are independent and identically distributed random errors with zero mean and residual variance $\sigma^2>0$.
	
	In nonparametric regression, the estimation of the mean function $g$ is a fundamental and important problem.
	An accurate estimate of the mean function is required for various purposes including, but not limited to, describing the relationship between responses and covariates, predicting observations at new experimental points, and imputing missing data values.
	In the existing literature, a large amount of effort has been made to obtain a reasonable estimate of $g$ by, e.g., the kernel method \citep{hardle1990applied,Wand1995bk}, the local linear method \citep{fan1996local}, and the smoothing spline method \citep{Wang2011bk}. Apart from the mean function, it is known that the residual variance also plays an important role and needs to be accurately estimated as well \citep{Dette1998}. For illustration, an estimate of the residual variance is needed for the purpose of constructing the confidence band for the mean function \citep{Rice1984}, testing the goodness of fit of the mean function \citep{Carroll1988bk,Gasser1991}, or assessing the discontinuities of the mean function \citep{Muller1999}.
	
	To estimate the residual variance, there are two main classes of methods in the literature: residual-based methods and difference-based methods.
	Residual-based methods are the classical approaches for estimating the residual variance, which often make use of the summation of squared residuals as follows:
	\begin{eqnarray}
		\tilde \sigma^2=\frac{1}{n-\nu}\sum_{i=1}^n\{Y_i-\tilde g(x_i)\}^2, \label{res_variance}
	\end{eqnarray}
	where $\tilde g$ is a nonparametric fit of the mean function and $\nu$ is the degrees of freedom associated with the fitted $\tilde g$.
	By \cite{HallMarron1990}, residual-based estimators are capable to achieve the asymptotically optimal rate for the mean squared error~(MSE) as
	$
	{\rm MSE}(\tilde \sigma^2)=n^{-1}{\rm var}(\varepsilon^2)+o(n^{-1}).
	$
	Despite the good theoretical properties, it is noteworthy that the practical performance of residual-based estimators may not necessarily be acceptable due to a couple of reasons.
	First, the performance of $\tilde \sigma^2$ in (\ref{res_variance}) is very sensitive to a delicate choice of the smoothing parameter.
	Second, the residual-based estimator $\tilde \sigma^2$ is a second-step estimator that directly follows from the mean function estimator $\tilde g$.
	In nonparametric regression, these two estimators are less likely to be independent of each other; and consequently, they may fail to provide a reliable confidence band for the mean function.
	
	In view of the above limitations, difference-based methods have emerged that provided popular alternatives for estimating the residual variance.
	They are often constructed as a linear combination of squared differences from the neighbouring observations, do not require a nonparametric fit of the mean function, and hence are easy to implement.
	For simplicity, we assume that the design points are equally spaced on $[0,1]$ with $x_i=i/n$ for $i=1,\dots,n$.
	For model (\ref{Model}), \cite{Rice1984} proposed the first-order difference-based estimator as
	$\hat \sigma^2_{\rm R}=\sum_{i=1}^{n-1}(Y_{i+1}-Y_i)^2/\{2(n-1)\}.$
	\cite{HallKayTit1990} further proposed the higher-order difference-based estimator as
	\begin{equation}
		\hat \sigma^2=\frac{1}{n-r}\sum_{i=1}^{n-r}\left(\sum_{j=0}^{r} d_jY_{i+j}\right)^2, \label{general form}
	\end{equation}
	where the difference sequence, $(d_0,\dots,d_r)$, is a sequence of real numbers satisfying
	\begin{equation}
		\sum_{j=0}^r d_j=0 \quad {\rm and} \quad \sum_{j=0}^r d_j^2=1 \label{C0}
	\end{equation}
	with $d_0d_r\neq 0$ and $d_0>0$, and $r$ is the order of the difference sequence.
	%Substituting the difference sequence in (\ref{general form}) results in various difference-based estimators.
	When $r=1$, the unique solution of the difference sequence under constraint~(\ref{C0}) is $(d_0,d_1)=(2^{-1/2},-2^{-1/2})$, which results in the Rice estimator.
	
	When $r\ge 2$, however, there are infinitely many solutions for the difference sequence under  constraint~(\ref{C0}).
	A natural question is then: which difference sequence is the best for estimator (\ref{general form}) to estimate the residual variance? 
	To answer this question, there are two popular difference sequences available in the literature: {\em the optimal difference sequence} and {\em the ordinary difference sequence}.
	The optimal difference sequence is to minimize the asymptotic variance of estimator (\ref{general form}), whereas the ordinary difference sequence is to eliminate the estimation bias up to order $r-1$. In other words, none of the two difference sequences took into account the bias-variance trade-off for the variance estimation.
	For the special case of $r=2$, \cite{HallKayTit1990} derived the optimal difference sequence as $(d_0,d_1,d_2) = (0.809,-0.5,-0.309)$ so that the resulting estimator is
	$\hat\sigma^2_{\rm HKT} = \sum_{i=1}^{n-2}(0.809Y_i - 0.5Y_{i+1} - 0.309Y_{i+2})^2/(n-2)$, and \cite{Gasser1986} derived the ordinary difference sequence as $(d_0,d_1,d_2) = (6^{-1/2}, -(2/3)^{1/2}, 6^{-1/2})$ so that the resulting estimator is
	$\hat\sigma^2_{\rm GSJ} = \sum_{i=1}^{n-2}(Y_i - 2Y_{i+1} + Y_{i+2})^2/\{6(n-2)\}$.

	For further discussion on the two difference sequences, \cite{Dette1998} conducted a comparative study and concluded that the ordinary difference sequence should be recommended when the sample size is small and the signal-to-noise ratio is large; while for other situations, the optimal difference sequence can be the default choice for practical use.
	Although very simple to implement, the application of their rule is somewhat limited since the signal-to-noise ratio is rarely known in practice.
	As a consequence, the choice of the difference sequence remains, in fact, rather arbitrary in the subsequent literature.
	Inspired by this,
	\cite{dai2017choice} proposed a unified framework for the variance estimation that
	combines the linear regression method with the higher-order difference-based estimators systematically.
	They further showed that, under the unified framework, the ordinary difference sequence can be consistently applied between the two difference sequences.

	In this paper, we propose to further advance the difference sequence selection from another unique perspective.
	To achieve this, we first reformulate the existing difference sequences as solutions to an optimization problem that minimizes the variance of the estimator under certain constraints. Then under the optimization framework, we propose a new family of difference sequence, called {\em the optimal-$k$ difference sequence}, by providing more flexible constraints on the estimation bias and variance.
	Moreover, we show that our newly proposed sequence is capable to achieve a better bias-variance trade-off in estimating the residual variance, and it also includes the optimal and ordinary difference sequences as two important special cases.
	Through theoretical and numerical studies, we demonstrate that the optimal-$k$ difference sequence has been pushing the boundaries of our knowledge in difference-based methods in nonparametric regression, and more importantly, it always performs the best in practical situations.

	The rest of this paper is organized as follows. In Section 2, we review the two existing difference sequences under the optimization framework for estimating the residual variance.
	In Section 3, we define the optimal-$k$ difference sequence, investigate its properties, and provide a procedure for generating the new sequence.
	In Section 4, we apply the optimal-$k$ difference sequence and introduce a new difference-based estimator called the optimal-$k$ estimator. We further study its asymptotic properties and draw connections with other difference-based estimators.
	In Section 5, we conduct simulation studies to evaluate the finite sample performance of the new estimator and compare it with the existing methods.
	Finally, we conclude the paper with some discussion and future work in Section 6, and provide the technical details in the Appendix.

	\section{Optimal and ordinary difference sequences}
	
	%For the difference-based estimator (\ref{general form}), we have
	%\begin{eqnarray*}
	%{\rm bias}(\hat \sigma^2)&=&{\rm E}(\hat \sigma^2)-\sigma^2={\frac{1}{n-r}}\sum_{i=1}^{n-r}(\sum_{j=0}^{r} d_jg_{i+j})^2, \\
	%{\rm var}(\hat \sigma^2)&=&{\frac{1}{n}}\{{\rm var}(\varepsilon^2)+4\delta\sigma^4\}+o(\frac{1}{n})
	%\end{eqnarray*}
	%\subsection{Difference-based estimator}
	\noindent
	To take a further look at the existing difference-based estimators, we first represent estimator (\ref{general form}) as a quadratic form of the observations as follows:
	\begin{equation}
		\hat \sigma^2=\frac{1}{n-r}Y^{ T}DY, \label{quadratic form}
	\end{equation}
	where $Y=(Y_1,\dots,Y_n)^T$ with $T$ denoting the transpose of a vector or a matrix, and $D={\tilde D}^{T}{\tilde D}$ with $\tilde D$ being an $(n-r)\times n$ matrix of the structure
	\begin{eqnarray*}
		\tilde D=\left(
		\begin{array}{cccccc}
			d_{0}&\dots &d_r &0&\dots &0   \\
			& \ddots & & \ddots &&\\
			& &\ddots  && \ddots &\\
			0& \ddots &0 & d_0 &\dots &d_r\\
		\end{array}
		\right).
	\end{eqnarray*}
	
	According to \cite{Dette1998}, the MSE of $\hat \sigma^2$ can be expressed as
	\begin{align}\label{quadratic mse}
		{\rm MSE}(\hat \sigma^2)=&~\frac{1}{(n-r)^2}\left[(g^{T}Dg)^2+4\sigma^2g^{T}D^2g+4g^{T}\{D{\rm diag}(D)u\}\sigma^3\gamma_3\right.\nonumber \\
		&\left.+~\sigma^4{\rm tr}\{{\rm diag}(D)^2\}(\gamma_4-3)+2\sigma^4{\rm tr}(D^2)\right],
	\end{align}
	where $g=(g(x_1),\dots,g(x_n))^{T}$, ${\rm diag}(D)$ denotes the diagonal matrix of the matrix $D$, $u=(1,\dots,1)^{T}$, $\gamma_i=E\{(\varepsilon/\sigma)^i\}$ for $i=3$ and $4$, and ${\rm tr}(D)$ denotes the trace of the matrix $D$.
	The first term in the right side of (\ref{quadratic mse}) is the squared bias, and the remaining four terms make up the variance of the estimator. When the random errors follow a normal distribution, the third and fourth terms in the right side of (\ref{quadratic mse}) will be zero so that the variance of the estimator can be further simplified as
	\begin{eqnarray*}
		{\rm var}(\hat \sigma^2)=\frac{1}{(n-r)^2}\left\{4\sigma^2g^{T}D^2g+2\sigma^4{\rm tr}(D^2)\right\}.
		\label{normal quadratic mse}
	\end{eqnarray*}
	
	In what follows, we review in detail the optimal and ordinary difference sequences. Also from a unique perspective, we reformulate them as solutions to an optimization problem that minimizes the MSE of the estimator under different constraints.

	%For an estimator with the form (\ref{general form}), the asymptotic estimation variance is
	%\begin{equation}
	%{\rm var}(\hat \sigma^2)={1\over n}\{{\rm var}(\varepsilon^2)+4\delta\sigma^4\}+o({1\over n}), \label{var-d}
	%\end{equation}
	%and the estimation bias is expressed as
	%\begin{equation}
	%{\rm bias}(\hat \sigma^2)={1\over n-r}\sum_{i=1}^{n-r}\left[\sum_{p=1}^r C_pg^{(p)}_i/(p!n^p)+o\left({1\over n^r}\right)\right]^2 .\label{bias-d}
	%\end{equation}
	%

	\subsection{Optimal difference sequence}
	
	\noindent
	Under some mild conditions, \cite{HallKayTit1990} showed that the estimation bias of estimator (\ref{quadratic form}) is asymptotically negligible compared to the estimation variance.
	They further derived the asymptotic MSE, or equivalently the asymptotic variance, of estimator (\ref{quadratic form}) as
	\begin{equation*}
		{\rm MSE}(\hat \sigma^2)={\frac{1}{n}}\{{\rm var}(\varepsilon^2)+4\sigma^4\delta(r)\}+o(\frac{1}{n}), \label{var-d}
	\end{equation*}
	where $\delta(r)=\sum_{c=1}^r(\sum_{j=0}^{r-c} d_jd_{j+c})^2$ for $r\ge 1$. It is clear that, besides the error moments and the sample size, the asymptotic MSE also depends on the choice of difference sequence $d(r)=(d_0,\dots, d_r)$ through $\delta(r)$.
	
	The optimal difference sequence, denoted by $d_{\text{opt}}(r)$, was defined as the minimizer of the asymptotic MSE. This is equivalent to minimizing the quantity $\delta(r)$ under the following optimization problem:
	\begin{equation}
		\underset{{d(r)\in \mathbb{R}^{r+1}}}{\rm arg~min}~\delta(r) \quad{\rm subject~to}\quad \sum_{j=0}^r d_j=0 ~{\rm and}~\sum_{j=0}^r d_j^2=1.\label{opt}
	\end{equation}
	To solve (\ref{opt}), the Lagrange multiplier can be readily applied so that the optimal difference sequence satisfies
	\begin{equation*}
		\sum_{j=0}^{r-c}d_{{\text{opt}},j}d_{{\text{opt}},j+c}=-\frac{1}{2r}, \quad 1\le c\le r. %\label{optimal_property}
	\end{equation*}
	They further lead to
	$\delta_{{\text{opt}}}(r) = \sum_{c=1}^r \{-1/(2r)\}^2 = 1/(4r)$, which is the minimum value of $\delta(r)$ associated with any difference sequence $d(r)$.
	
	We refer to estimator (\ref{quadratic form}) with the optimal difference sequence as the optimal estimator, denoted by $\hat \sigma^2_{\text{opt}}(r)$. When the design points are equally spaced and the mean function has a bounded first derivative, it can be further shown that
	\begin{eqnarray*}
		{\rm var}\{\hat \sigma^2_{\text{opt}}(r)\}={\frac{1}{n}}\{{\rm var}(\varepsilon^2)+4\sigma^4\delta_{{\text{opt}}}(r) \}+o({\frac{1}{n}})\quad {\rm and}\quad {\rm bias}\{\hat \sigma^2_{\text{opt}}(r)\}=O({\frac{1}{n^{2}}}).
	\end{eqnarray*}
	This coincides with the result in \cite{HallKayTit1990} that the estimation bias is asymptotically negligible compared to the estimation variance.

	%Consequently, $\delta_{\text{opt}}=1/4r$ and hence ${\rm var}\{\hat \sigma^2_{\text{opt}}(r)\}\rightarrow\{{\rm var}(\varepsilon^2)+\sigma^4/r\}/n,$
	%where $\hat \sigma^2_{\text{opt}}(r)$ denotes the estimator constructed with the order-$r$ optimal sequence.

	\subsection{Ordinary difference sequence}
	\noindent
	When the sample size is small and the mean function is rough, it is known that the estimation bias of $\hat \sigma_{\text{opt}}^2$ is no longer negligible, or more seriously, it may even dominante the MSE. For the variance estimation in such scenarios, the ordinary difference sequence was then introduced that aims to eliminate the estimation bias as much as possible \citep{Gasser1986,Buckley1988,Seifert1993}.
	%Let $d_{\text{ord}}(r)$ denote the ordinary difference sequence.
	%To get the order-$r$ ordinary difference sequence, we can simply follow the procedure below:
	%\begin{itemize}
	%	\item [(1)] Define $y_i^{(1)}=(y_{i+1}-y_{i})/(x_{i+1}-x_i)$.
	%	\item [(2)] Inductively define $y_i^{(r)}=(y_{i+1}^{(r-1)}-y_{i}^{(r-1)})/(x_{i+r}-x_i)$, $r=2,3,\dots$.
	%	\item [(3)] Write $y_i^{(r)}$ as a linear combination of $y_i,\dots,y_{i+r}$, i.e., $y_i^{(r)}=\sum_{j=0}^rw_{ij}y_{i+j}$.
	%	\item [(4)] The order-$r$ ordinary difference sequence: $(d_{i0},\dots,d_{ir})=(w_{i0},\dots,w_{ir})/\sqrt{\sum_{j=0}^rw_{ij}^2}$.
	%\end{itemize}
	
	%, denoted by $d_{\text{ord}}(r)$, 
	
	By (\ref{quadratic mse}), the bias term is given as
	\begin{equation*}
		{\rm bias}(\hat \sigma^2)=\frac{g^{T}Dg}{n-r}={\frac{1}{n-r}}\sum_{i=1}^{n-r}\left(\sum_{j=0}^{r} d_jg(x_{i+j})\right)^2.
	\end{equation*}
	We further assume that the mean function has a bounded $r$th derivative. Then under the equidistant design, by the Taylor expansion it follows that
	\begin{equation}\label{expansion}
		\sum_{j=0}^{r} d_jg(x_{i+j})=\sum_{j=0}^{r} d_j\left\{\sum_{p=0}^r \frac{j^p}{p!n^p}g^{(p)}(x_i)+o({\frac{1}{n^{r}}})\right\}=\sum_{p=0}^r \frac{C_p}{n^p}g^{(p)}(x_i)+o({\frac{1}{n^{r}}}),
	\end{equation}
	where $g^{(p)}$ denotes the $p$th derivative and $C_p=\sum_{j=0}^r j^{p}d_j/p!$ for $p=0,1,\ldots,r$.
	Finally, by plugging the approximate terms in (\ref{expansion}) back to the bias formula, we have
	\begin{equation}
		{\rm bias}(\hat \sigma^2)={\frac{1}{n-r}}\sum_{i=1}^{n-r}\left\{\sum_{p=0}^r \frac{C_p}{n^p}g^{(p)}(x_i)+o({\frac{1}{n^{r}}})\right\}^2 .\label{bias-d}
	\end{equation}
	
	Noting that the difference sequence $d(r)$ consists of $r+1$ unknown quantities, one can (and only can) impose a maximum of $r+1$ constraints on the difference sequence for the purpose of eliminating the estimation bias. 
	To be more specific, apart from the two minimum requirements $C_0=\sum_{j=0}^r d_j=0$ and $\sum_{j=0}^r d_j^2=1$ in (\ref{C0}) for model identifiability, we can impose a maximum of $r-1$ additional constraints $C_1=\cdots=C_{r-1}=0$ to further eliminate the bias term in (\ref{bias-d}). 
	To conclude, the ordinary difference sequence can be redefined as the minimizer of the following optimization problem: 
	%To minimize the order of the estimation bias, the difference sequence needs to satisfy
	%\begin{eqnarray}
	%\sum_{j=0}^rd_j^2=1~ {\rm and}~ C_0=\cdots=C_{r-1}=0. \label{C_r}
	%\end{eqnarray}
	%Consequently, we have
	%\begin{eqnarray*}
	%{\rm var}\{\hat \sigma^2_{\text{ord}}(r)\}&=&{1\over n}\{{\rm var}(\varepsilon^2)+\sigma^4\delta(d_{\text{ord}}(r))\}+o(n^{-1}),\\
	%{\rm bias}\{\hat \sigma^2_{\text{ord}}(r)\}&=&C_{r}^2J_{r}+o\left({n^{-2r}}\right),
	%\end{eqnarray*}
	%where $\hat \sigma^2_{\text{ord}}(r)$ denotes the estimator constructed with the order-$r$ optimal difference sequence and
	%$J_{k}=\int_0^1[g^{(k)}(x)]^2dx$, $k=0,\dots,r$.
	%Under the optimization framework, we can treat $d_{\text{ord}}(r)$ as the solution to the following problem:
	\begin{equation}
		\underset{{d(r)\in \mathbb{R}^{r+1}}}{\rm arg~min}~\delta(r)\quad{\rm subject~to}\quad C_0=C_1=\cdots=C_{r-1}=0~ {\rm and}~ \sum_{j=0}^rd_j^2=1. \label{ord}
	\end{equation}
	It is also interesting to point out that, by solving the $r+1$ constraints, it yields a unique solution of the ordinary difference sequence as $d_{\rm ord}(r)=(d_{{\text{ord}},0},\dots,d_{{\text{ord}},r})$ where 
	\begin{equation}
		d_{{\text{ord}},j}=(-1)^j {r \choose j}{2r \choose r}^{-1/2},\quad j=0,\dots,r. \label{ordinary}
	\end{equation}
	In other words, the optimization problem in (\ref{ord}) is, in fact, a degenerate optimization problem and the minimum value of $\delta(r)$ is fixed as $\delta(d_{\rm ord}(r))$.

	Moreover, with the ordinary difference sequence in (\ref{ordinary}), we refer to estimator (\ref{quadratic form}) as the ordinary estimator, denoted by $\hat \sigma^2_{\text{ord}}(r)$.
	\cite{Dette1998} also derived that
	\begin{eqnarray*}
		{\rm var}\{\hat \sigma^2_{\text{ord}}(r)\}={\frac{1}{n}}\left\{{\rm var}(\varepsilon^2)+4\sigma^4\delta_{{\text{ord}}}(r)\right\}+o({1\over n})\quad {\rm and} \quad {\rm bias}\{\hat \sigma^2_{\text{ord}}(r)\}=O({1\over n^{2r}}),
	\end{eqnarray*}
	where $\delta_{{\text{ord}}}(r)=[{2r \choose r}^{-2}{4r \choose 2r}-1]/2$.
	As expected, the ordinary estimator can control the estimation bias up to order $O({n^{-2r}})$, compared to the minimum control of $O(n^{-2})$ for the optimal estimator.
	As a trade-off, however, the ordinary estimator has a larger asymptotic variance than the optimal estimator, especially when the order $r$ is high.

	\section{Optimal-$k$ difference sequence}

	\subsection{Definition}
	\begin{figure}[!b]
		\centering
		\includegraphics[width=13cm]{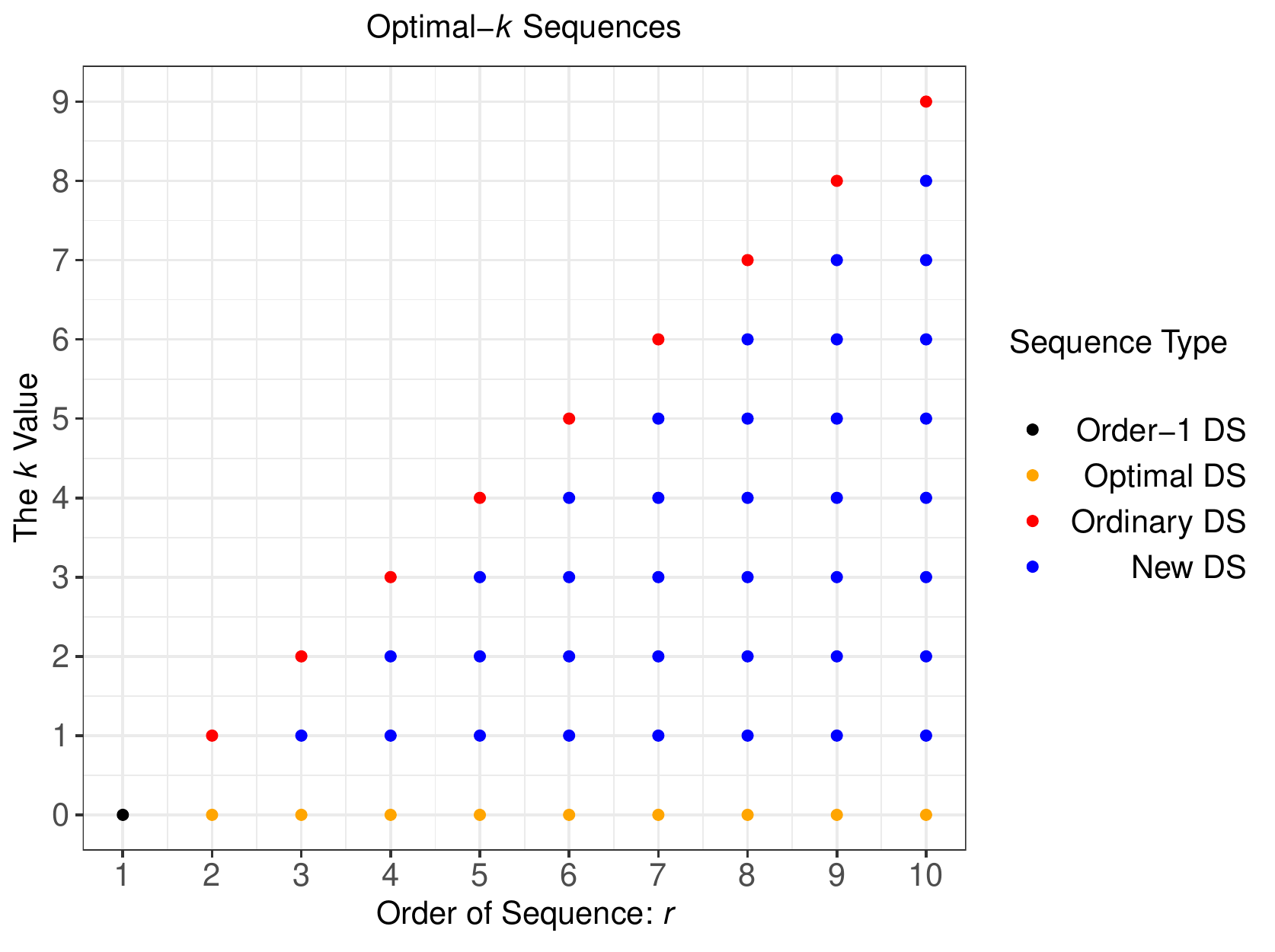}\\
		\caption{The unified framework of optimal-$k$ difference sequences, where the existing difference sequences (DS) are all located on the boundary of the triangle region as important special cases.}
		\label{framework}
	\end{figure}
	\noindent
	As shown in Section 2, the optimal and ordinary difference sequences can be derived as the unique solutions under the optimization problems (\ref{opt}) and (\ref{ord}), respectively. Specifically, 
	the first one provides a minimum control on the estimation bias, and the second one may over-control the estimation bias so that, as a price to pay, the estimation variance is dramatically enlarged. 
	To overcome the limitations on the two extreme cases, we propose a compromise solution that defines the difference sequence as the solution(s) to the following unified optimization problem:
	\begin{equation}
		\begin{gathered}
			\underset{{d(r)\in \mathbb{R}^{r+1}}}{\rm arg~min}~\delta(r)=\sum_{c=1}^r\left(\sum_{j=0}^{r-c} d_jd_{j+c}\right)^2  \\
			{\rm subject~to}\quad  C_0=C_1=\cdots=C_{k}=0\quad {\rm and}\quad \sum_{j=0}^rd_j^2=1,
		\end{gathered} \label{opt-k}
	\end{equation}
	where $0\le k \le r-1$ is an integer number. In the special case when $k=0$, our unified optimization problem reduces to (\ref{opt}); and in the special case when $k=r-1$, it reduces to (\ref{ord}). 
	When $k$ increases from 0 to $r-1$, with the increased number of constraints one can control the estimation bias at a higher order; yet on the other side, with a smaller space for $d(r)$, the minimum possible variance of estimator (\ref{quadratic form}) will be enlarged as a trade-off.

	To further explore the effect of $k$ on the optimization estimation, we also present the difference sequences in Figure \ref{framework} for various $r$ values. 
	When $r=1$, there is only one unique sequence, represented by the black point in the corner, known also as the Rice difference sequence. When $r=2$, there are two options of difference sequences, one is optimal and the other is ordinary. When $r\ge 3$, the optimization problem will produce a new difference sequence for each $1\le k\le r-2$. For ease of presentation, we refer to it as \emph{the optimal-$k$ difference sequence} and denote it by ${d}_{k}(r)$. To summarize, the orange points with $k=0$ represent the optimal difference sequences, the red points on the diagonal line $k=r-1$ represent the ordinary difference sequences, and the blue points in the middle part represent the newly introduced optimal-$k$ difference sequences. 
	
	%In view of this, we have greatly enlarged the family of difference sequences, i.e., from two boundary lines to a sector of the plane.

	%
	%By comparing the three problems (\ref{opt}), (\ref{ord}) and (\ref{opt-k}), we see that the existing difference sequences are special cases of the optimal-$k$ difference sequence.
	%For instance, when $r=1$, $k=0$ is the only possible case and the unique solution is $(1/\sqrt{2},-1/\sqrt{2})$ which was used in \cite{Rice1984}.
	%When $r=2$ and $k=0$, the solution is $d_{0,2}=d_{\text{opt}}(2)=(0.809,-0.5,-0.309)$ in \cite{HallKayTit1990}.
	%When $r=2$ and $k=1$, the solution is $d_{1,2}=d_{\text{ord}}(2)=(6^{-1/2}, -(2/3)^{1/2}, 6^{-1/2})$ in \cite{Gasser1986}.
	%When $r\ge 3$ and $k=0$ or $r-1$, $d_{0,r}=d_{\text{opt}}(r)$ and $d_{r-1,r}=d_{\text{ord}}(r)$; when $1\le k\le r-2$, $d_{k,r}$ are new difference sequences that have not been investigated in the literature.
	%The above correspondence is intuitively demonstrated by the lower triangle region in Figure \ref{framework}.
	%Apparently, when $k\ge r$, no valid difference sequences exist so the upper triangle region is empty.
	%The black point corresponds to the Rice's difference sequence; the cyan points on the line of $k=0$ correspond to the optimal difference sequences; the red points on the diagonal correspond to the ordinary difference sequences; the blue points in between correspond to the new members of the optimal-$k$ difference sequence. In view of this, we have greatly enlarged the family of difference sequences, i.e., from two boundary lines to a half plane.
	
	\subsection{Properties and generating procedure}
	
	\noindent
	In this section, we propose to minimize the objective function, $\delta(r)$, in (\ref{opt-k}) and provide an algorithm for generating ${d}_{k}(r)$.
	To start with, we let $I_k=\sum_{c=1}^r c^k$ for $k=0,1, \dots$, and 
	\begin{equation}
		{V}_{k}=\left(
		\begin{array}{cccc}
			I_0 & I_2  & \cdots & I_{2k}\\
			I_2 & I_4  & \cdots & I_{2k+2}\\
			\vdots & \vdots & \ddots &\vdots \\
			I_{2k} & I_{2k+2}& \cdots& I_{4k}\\
		\end{array}
		\right). \label{V_k}
	\end{equation}
	In the following theorem, we first show that $V_k$ is nonsingular and then derive the minimum value of $\delta(r)$ in closed form for arbitrary optimal-$k$ difference sequence. 
	
	\begin{theorem}\label{delta-kr}
		(a) For any $0\le k \le r-1$, the constant matrix $V_k$ is an invertible matrix. (b)~Let ${V}_{k}^{-1}$ be the inverse matrix of ${V}_{k}$ and ${V}_{k}^{-1}(i,j)$ be the element of ${V}_{k}^{-1}$ on the $i$th row and $j$th column. Then for any given $r\geq 1$, the minimum value of $\delta(r)$ under the optimization problem (\ref{opt-k}) with fixed $k$ is 
		$$\delta_k(r)=\frac{1}{4}{V}_{k}^{-1}(1,1).$$
		%(a) For any $0\le k \le r-1$, the constant matrix $V_k$ in (\ref{V_k}) is an invertible matrix. (b) For any fixed $r\geq 1$, the minimum value of $\delta(r)$ under the optimization problem (\ref{opt-k}) with $0\leq k\leq r-1$ is 
		%$$\delta_k(r)=\frac{1}{4}{V}_{k}^{-1}(1,1).$$
	\end{theorem}

	%\begin{figure}[!b]
	%	\centering
	%	% Requires \usepackage{graphicx}
	%	\includegraphics[width=\textwidth]{delta-map}\\
	%	\caption{The values of $\delta_k(r)$. The red curve: $d_{0}(r)$; the blue curve: $d_{1}(r)$; the yellow curve: $d_{2}(r)$; the purple curve: $d_{3}(r)$; the green curve: $d_{4}(r)$; the black curve: $d_{r-1}(r)$.}
	%	\label{delta-map}
	%\end{figure}
	
	\vskip 12pt
	The proof of Theorem 1 is given in the Appendix A. 
	By Theorem 1 and the fact that ${V}_{k}$ is a constant matrix for any given pair $(r,k)$, we can derive $\delta_k(r)$ for any optimal-$k$ difference sequence.
	When $k=0$, we have $V_0=I_0=r$ and so $\delta_0(r)=1/(4r)$, which is the same as $\delta_{\text{opt}}(r)$ in \cite{HallKayTit1990}; see also Section 2.2 for more details.
	When $k=r-1\ge 1$, it results in the ordinary difference sequence
	$$\delta_{r-1}(r)=\frac{1}{4}{V}_{r-1}^{-1}(1,1)=\frac{1}{2}\left\{{2r \choose r}^{-2}{4r \choose 2r}-1\right\},$$ 
	which also coincides with the results in \cite{Dette1998}. 
	While for the middle $k$ values such that $1\leq k\leq r-2$ with $r\ge 3$, we let $\mathbb{S}_{r,k}$ denote the feasible region of (\ref{opt-k}), i.e., the set of difference sequences satisfying the respective constraints.
	Then by noting that the constraints in (\ref{opt}), (\ref{ord}) and (\ref{opt-k}) are nested, we have $\mathbb{S}_{r,r-1} \subset \mathbb{S}_{r,k} \subset \mathbb{S}_{r,0}$, and consequently, $\delta_0(r)\le \delta_k(r)\le \delta_{r-1}(r)$. 
	%To further assess the behavior of $\delta_k(r)$ against $k$ and $r$, we calculate $\delta_k(r)$ for $\{(3,r): 0\le k \le {\rm min}(4,r-1),~ 1\le r \le 15\}$ by Theorem 1. %and plot the numerical results in Figure \ref{delta-map}.
	According to Theorem 1, we calculate $\delta_k(r)$ for a range of difference sequences and find that $\delta_k(r)$ increases with $k$ for a fixed $r$, i.e., $\delta_k(r)< \delta_{k+1}(r)$ for $1\le k \le r-2$, which coincides with the previous analytical results.
	For a fixed $k$, $\delta_k(r)$ monotonically decreases to zero as $r$ is getting larger. Exceptionally, $\delta_{r-1}(r)$ monotonically increases with $r$ for the ordinary difference sequence.

	%This is because, for a fixed $k$ and $\tilde r> k$, $\mathbb{S}_{k,\tilde r}\subset \mathbb{S}_{k, \tilde r+1}$: increasing the order of sequence allows us to search for the minima of $\delta(r)$ in a broader space.
	%$\delta_{r-1}(r)$ increases with $r$ because once we increase the order of sequence, the number of constraints in (\ref{ord}) also increases accordingly; as a result, the feasible space is always a single point for the problem (\ref{ord}).
	%Additionally, the curves of $\delta_{k}(r)$, $1\le k\le 4$, decease to zero as $r$ gets large enough. 
	%In particular, according to Theorem 1, we may derive that
	%$$\frac{\delta_{k}(r)}{\delta_{0}(r)}\rightarrow \tilde{ {V}}_{k} ^{-1}(1,1)\quad {\rm as}\quad r\to \infty,$$
	%where $\tilde {{V}}_{k} $ is a $(k+1)\times (k+1)$ matrix and $\tilde {{V}}_{k} (i,j)=1/\{2(i+j)-3\}$.

	%To implement the sequences, we also provide a practical procedure to generate the optimal-$k$ difference sequence. 
	%Other than the derivation of minimum value for $\delta_{k,r}$, we also provide a practical procedure to generate the optimal-$k$ difference sequence. 
	Except for the ordinary difference sequence in (\ref{ordinary}), a closed-form solution may not exist for most difference sequences in the optimal-$k$ family. 
	In the following theorem, we show that, for any given pair $(r,k)$,  the optimal-$k$ difference sequence from the optimization problem (\ref{opt-k}) can be alternatively derived as the solution to a root-finding problem.

	\begin{theorem}\label{calculation}
		Let $R(t)=t^r\{D_r(t^r+t^{-r})+\cdots+D_1(t+t^{-1})+1\}$ be a self-reciprocal polynomial, 
		where $D_c=\sum_{j=0}^{r-c}d_jd_{j+c}=-\sum_{s=0}^kc^{2s}{V}_{k}^{-1}(s+1,1)/2~{\rm for}~c=1,\dots,r.$ Let also $\{1,z_2,\dots,z_r,1,z_2^{-1},\dots,z_r^{-1}\}$ be the $2r$ roots from the equation $R(t)=0$. 
		We further choose one unit root and the $r-1$ roots 
		outside the unit circle, denoted by $\{t_1,\dots,t_r\}$, and apply them to construct a new polynomial as
		$$(x-t_1)(x-t_2)\cdots(x-t_{r})=\sum_{k=0}^r a_kx^kz,$$
		where $a_k$ are the coefficients. Then the optimal-$k$ difference sequence is given as
		$$(d_0,\dots,d_r)=(a_0,\dots,a_r)/(\sum_{i=0}^r a_i^2)^{1/2}.$$
		%where $\|\cdot\|_2$ denotes the $L_2$ norm of a vector.
	\end{theorem}
	
		\begin{table}[!b]
		%\small
		\caption{Optimal-$k$ difference sequences $d_k(r) = (d_0, \dots, d_r)$ for $r\leq 5$, where $d_0>0$ for identifiability and the entries are rounded to four decimal places.}
		\label{opt2}
		\begin{center}
			\begin{tabular}{c|cccccc} \hline
				% &\multicolumn{7}{c}{$(d_0,...,d_r)$}  \\ \hline\hline
				$(r,k)$ &$d_0$&$d_1$& $d_2$& $d_3$& $d_4$& $d_5$   \\\hline \hline
				(1,~0) &0.7071& -0.7071& &&&\\\hline
				(2,~0) &0.8090& -0.5000& -0.3090&&&\\
				(2,~1) &0.4082 &-0.8165 & ~0.4082&&&\\\hline
				(3,~0) &0.8582& -0.3832&-0.2809&-0.1942&&\\ 
				(3,~1) &\textbf{0.2673}& ~\textbf{0.0000}&\textbf{-0.8018}&\textbf{~0.5345}& &\\ 			
				(3,~2) &0.2236&-0.6708&~0.6708 &-0.2236& &\\ \hline
				(4,~0) &0.8873& -0.3099&-0.2464&-0.1901& -0.1409&\\ 
				(4,~1) &\textbf{0.1982}&  ~\textbf{0.1034}&\textbf{-0.1855}&\textbf{-0.7322}&~\textbf{0.6160}&\\ 		(4,~2) &\textbf{0.1842}& \textbf{-0.2271}&\textbf{-0.4242}&~\textbf{0.7928}& \textbf{-0.3257}&\\ 
				(4,~3) &0.1195& -0.4781&~0.7171&-0.4781&~0.1195&\\ \hline
				(5,~0) &0.9064& -0.2600&-0.2167&-0.1774&-0.1420&-0.1103\\ 
				(5,~1) &\textbf{0.1573}&  ~\textbf{0.1151}&\textbf{-0.0166}&\textbf{-0.2681}&\textbf{-0.6609}&~\textbf{0.6732}\\ 			
				(5,~2) &\textbf{0.1553} &\textbf{-0.0734}& \textbf{-0.3080} &\textbf{-0.1904} &~\textbf{0.8217}& \textbf{-0.4053}\\ 
				(5,~3) &\textbf{0.1143} &\textbf{-0.2726} &\textbf{-0.0603} &~\textbf{0.6733} &\textbf{-0.6470}  &~\textbf{0.1923}\\ 
				(5,~4) &0.0630 &-0.3150  &~0.6299 &-0.6299  &~0.3150 &-0.0630\\  \hline 
		\end{tabular}\end{center}
	\end{table}

	\vskip 12pt
	The proof of Theorem 2 is given in the Appendix. By Theorem 2, for any given pair $(r,k)$, all the values of $D_c$ and ${V}_{k}$ can be explicitly computed, and consequently, the roots of $R(t)=0$ can be solved with some classical algorithms for root-finding. Note that there are different ways to choose the $r$ roots in Theorem 2. In other words, the optimal-$k$ difference sequence may not be unique for any fixed $(r,k)$, which, in fact, was also shown by \cite{yatchew2003semiparametric} that the uniqueness of the optimal difference sequence claimed in \cite{HallKayTit1990} is incorrect. In theory, however, different options of $d_k(r)$ will lead to equivalent estimators and so they can be treated indifferently. We thus apply the algorithm in Theorem \ref{calculation} to generate the optimal-$k$ difference sequence sequence for any part $(r,k)$ without loss of generality; while for easy reference, Table 1 also provides the numerical results of difference sequences for the order $r$ up to $5$.

	%\begin{table}[H]
	%\small
	%\caption{Optimal-$1$ sequences, $d_1(r)$, for $2\le r\le 6$. Entries are rounded to four decimal places.}
	%\label{opt1}
	%\begin{center}
	%\begin{tabular}{cccccccc} \hline
	% % &\multicolumn{7}{c}{$(d_0,...,d_r)$}  \\ \hline\hline
	% $r$ &$d_0$&$d_1$& $d_2$& $d_3$& $d_4$& $d_5$& $d_6$   \\\hline \hline
	% 2& 0.4082 &-0.8165 & 0.4082&&&&\\
	% 3&0.2673& 0.0000&-0.8018& 0.5345&&&\\
	% 4&0.1982& 0.1034&-0.1855&-0.7322&0.6160&&\\
	% 5&0.1573& 0.1151&-0.0166&-0.2681&-0.6609&0.6732&\\
	% 6&0.7155&-0.5976&-0.3030&-0.0934&0.0395&0.1088&0.1303\\ \hline
	%\end{tabular}\end{center}
	%\end{table}
	%
	%\begin{table}[H]
	%	\small
	%	\caption{Optimal-$2$ sequences, $d_2(r)$, for $3\le r\le 6$. Entries are rounded to four decimal places.}
	%	\label{opt2}
	%	\begin{center}
	%		\begin{tabular}{cccccccc} \hline
	%			% &\multicolumn{7}{c}{$(d_0,...,d_r)$}  \\ \hline\hline
	%			$r$ &$d_0$&$d_1$& $d_2$& $d_3$& $d_4$& $d_5$& $d_6$   \\\hline \hline
	%			3&0.2236& -0.6708& 0.6708&-0.2236&&&\\
	%			4&0.3257&-0.7927&0.4243& 0.2271&-0.1842&&\\
	%			5&0.4049& -0.8217&0.1909&0.3081& 0.0732&-0.1554&\\
	%			6&0.4680&-0.8118&0.0207&0.2628&0.1856&0.0088&-0.1340\\ \hline
	%	\end{tabular}\end{center}
	%\end{table}
	%
	%0.2673  0.0000 -0.8018  0.5345

	\section{Optimal-$k$ estimator}
	
	\noindent
	In this section, we apply the optimal-$k$ difference sequence in Section 3 to estimate the residual variance in model (\ref{Model}). We also refer to the new estimator as the optimal-$k$ estimator, denoted by $\hat \sigma^2_{k}(r)$ for $0\leq k \leq r-1$. Then as two special cases, the optimal and ordinary estimators are given as 
	$\hat \sigma^2_{0}(r)$ and $\hat \sigma^2_{r-1}(r)$, respectively.
	Now to study the effect of $(r,k)$ on the optimal estimation, we have the following theorem on the asymptotic bias and variance of the optimal-$k$ estimator, with the proof in the Appendix.

	%
	%
	%
	%To introduce the properties of the optimal-$k$ estimator, we first employ some notions as follows:
	%\begin{equation*}
	%{\rm V}_k\triangleq\left(
	%  \begin{array}{cccc}
	%    I_0 & I_2  & \cdots & I_{2k}\\
	%    I_2 & I_4  & \cdots & I_{2(k+1)}\\
	%    \vdots & \vdots & \ddots &\vdots \\
	%    I_{2k} & I_{2(k+1)}& \cdots& I_{4k}\\
	%  \end{array}
	%\right)\quad {\rm and}\quad I_k\triangleq\sum_{l=1}^r l^k.
	%\end{equation*}
	%We have omitted the dependence of $V_k$ and $I_k$ on the order of difference sequence, $r$, for simplicity.
	%Denote $V_k^{-1}$ as the inverse matrix of $V_k$, which is apparently invertible, and denote $V_k^{-1}(i,j)$ as the element of $V_k^{-1}$ on its $i$th row and $j$th column.
	\begin{theorem}\label{optkr}
		Assume that the mean function has a continuous $r$th derivative for any given $r\geq 1$ and $E(\varepsilon^4)<\infty$. Then under the equidistant design, we have
		\begin{eqnarray*}
			{\rm bias}\{\hat \sigma^2_{k}(r)\}&=&\frac{C_{k+1}^2}{n^{2(k+1)}}J_{k+1}+o({1\over n^{2(k+1)}}), \label{opt-k-bias} \\
			{\rm var}\{\hat \sigma^2_{k}(r)\}&=&{\frac{1}{n}}\left\{{\rm var}(\varepsilon^2)+\sigma^4{V}_{k}^{-1}(1,1)\right\}+o({1\over n}),\label{opt-k-var}
		\end{eqnarray*}
		where $C_k=\sum_{j=0}^r j^{k}d_j/k!$ and $J_{k}=\int_0^1\{g^{(k)}(x)\}^2dx$, $k=0,\dots,r-1$.
	\end{theorem}

	%\begin{figure}[!b]
	%	\centering
	%	% Requires \usepackage{graphicx}
	%	\includegraphics[width=\textwidth]{delta-map}\\
	%	\caption{The values of $\delta_k(r)$. The red curve: $d_{0}(r)$; the blue curve: $d_{1}(r)$; the yellow curve: $d_{2}(r)$; the purple curve: $d_{3}(r)$; the green curve: $d_{4}(r)$; the black curve: $d_{r-1}(r)$.}
	%	\label{delta-map}
	%\end{figure}
	
	\vskip 5pt
	%The proof of Theorem 3 is given in the Appendix.
	%Theorem 3 is an immediate result from Theorem 1, equation (9) and the fact that the asymptotic estimation variance of $\hat \sigma^2_{k}(r)$ is ${\frac{1}{n}}\{{\rm var}(\varepsilon^2)+4\sigma^4\delta(r)\}$ \cite{HallKayTit1990,Dette1998}. 
	Note that Theorem \ref{optkr} also includes the asymptotic results for the existing difference-based estimators. In the special case when $k=0$, we have the optimal estimator in \cite{HallKayTit1990}  with the same asymptotic results as 
	%$\hat \sigma^2_{\text{opt}}=\hat \sigma^2_{0}(r)$ and
	\begin{eqnarray*}
		{\rm bias}\{\hat \sigma^2_{0}(r)\}&=&\frac{C_{1}^2}{n^2}J_{1}+o({1\over n^{2}}),\\
		{\rm var}\{\hat \sigma^2_{0}(r)\}&=&{\frac{1}{n}}\left\{{\rm var}(\varepsilon^2)+\frac{1}{r}\sigma^4\right\}+o({1\over n}).
	\end{eqnarray*}
	On the other side, when $k=r-1$ with $r\ge 2$,  we have the ordinary estimator in \cite{Dette1998} with the same asymptotic results as 
	%$\hat \sigma^2_{\text{ord}}(r)=\hat \sigma^2_{r-1}(r)$ and
	\begin{eqnarray*}
		{\rm bias}\{\hat \sigma^2_{r-1}(r)\}&=&\frac{C_{r}^2}{n^{2r}}J_{r}+o({1\over n^{2r}}),\\
		{\rm var}\{\hat \sigma^2_{r-1}(r)\}&=&{\frac{1}{n}}\left\{{\rm var}(\varepsilon^2)+\left(2{2r \choose r}^{-2}{4r \choose 2r}-2\right)\sigma^4\right\}+o({1\over n}).
	\end{eqnarray*}
	Now for the new estimator, if we consider $k=1$ with $r\ge 3$, we have ${V}_{1}^{-1}(1,1)=4I_4/(I_0I_4-I_2^2)\rightarrow 9/(4r)$ as $r\rightarrow \infty$ and thus
	\begin{eqnarray*}
		{\rm bias}\{\hat \sigma^2_{1}(r)\}&=&\frac{C_{2}^2}{n^4}J_{2}+o({1\over n^{4}}),\\
		{\rm var}\{\hat \sigma^2_{1}(r)\}&=&{\frac{1}{n}}\left\{{\rm var}(\varepsilon^2)+{9\over 4r}\sigma^4\right\}+o({1\over n}).
	\end{eqnarray*}
	
	In what follows, we show that the optimal-$1$ estimator may provide a reasonable compromise between the optimal and ordinary estimators.
	For ease of explanation, we assume that $\{\varepsilon_i\}$ are normal errors so that ${\rm var}(\varepsilon^2)=2\sigma^4$. Then to compare the optimal-$1$ estimator and the optimal estimator in \cite{HallKayTit1990}, we note that ${\rm var}\{\hat \sigma^2_{1}(r)\}/{\rm var}\{\hat \sigma^2_{0}(r)\}=1+5/(8r+4)+o(1)$ and ${\rm bias}\{\hat \sigma^2_{1}(r)\}/{\rm bias}\{\hat \sigma^2_{0}(r)\}=(C_{2}^2J_{2})/(C_{1}^2J_{1})n^{-2}=O(n^{-2})$. To conclude, the optimal-$1$ estimator
	significantly reduces the asymptotic bias compared to the optimal estimator; yet as a price to pay, there is a little sacrifice in the asymptotic variance, which however can be small when $r$ is three or more. Next, for a comparison between the optimal-$1$ estimator and the ordinary estimator in \cite{Gasser1986}, we note that 
	${\rm var}\{\hat \sigma^2_{1}(r)\}/{\rm var}\{\hat \sigma^2_{r-1}(r)\}=(\pi r/2)^{-1/2}+o(r^{-1/2})$ and ${\rm bias}\{\hat \sigma^2_{1}(r)\}/{\rm bias}\{\hat \sigma^2_{r-1}(r)\}=O(n^{2(r-2)})$. To conclude, the optimal-$1$ estimator significantly reduces the asymptotic variance compared to the ordinary estimator; while for the asymptotic bias, given that the optimal-1 estimator has already controlled the order at $n^{-4}$, a further reduction may not bring in significant improvement in practice.  
		\begin{figure}[!t]
		\centering
		\begin{tabular}{cc}
			\includegraphics[width=7.5cm]{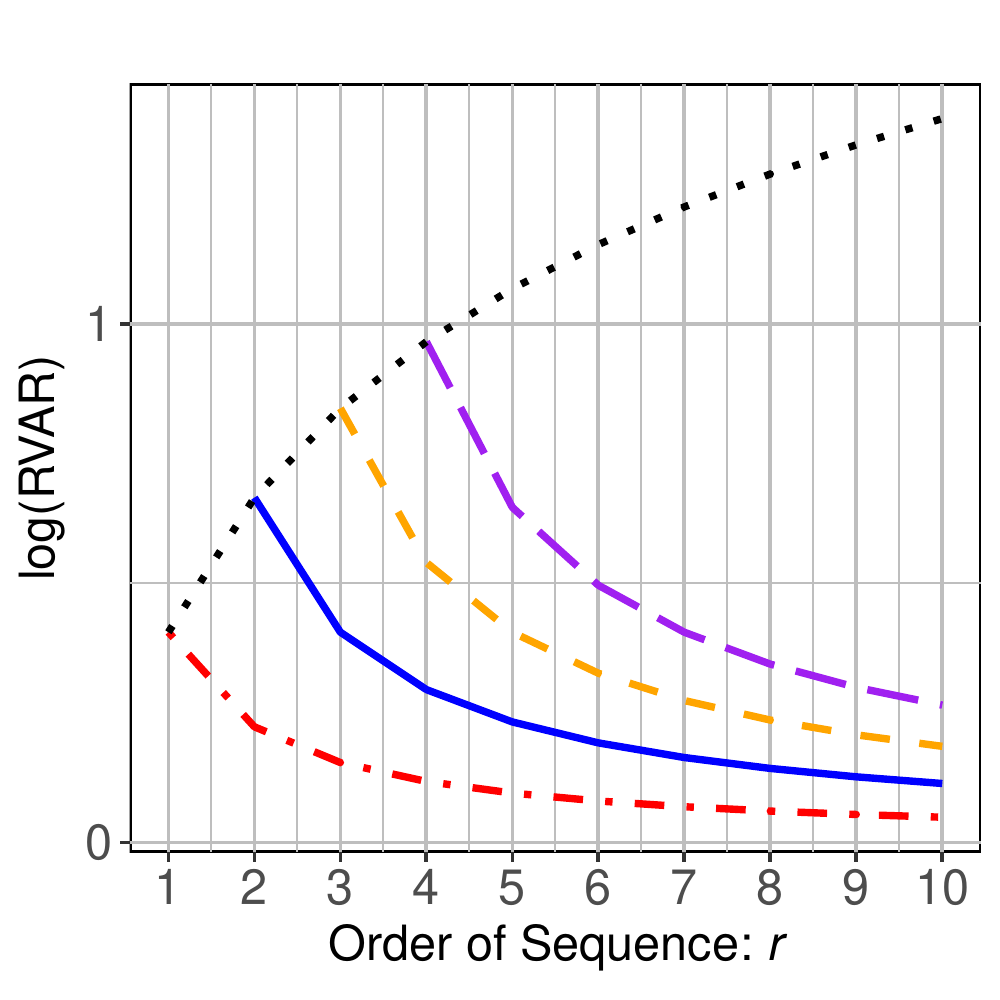} &   
			\includegraphics[width=7.5cm]{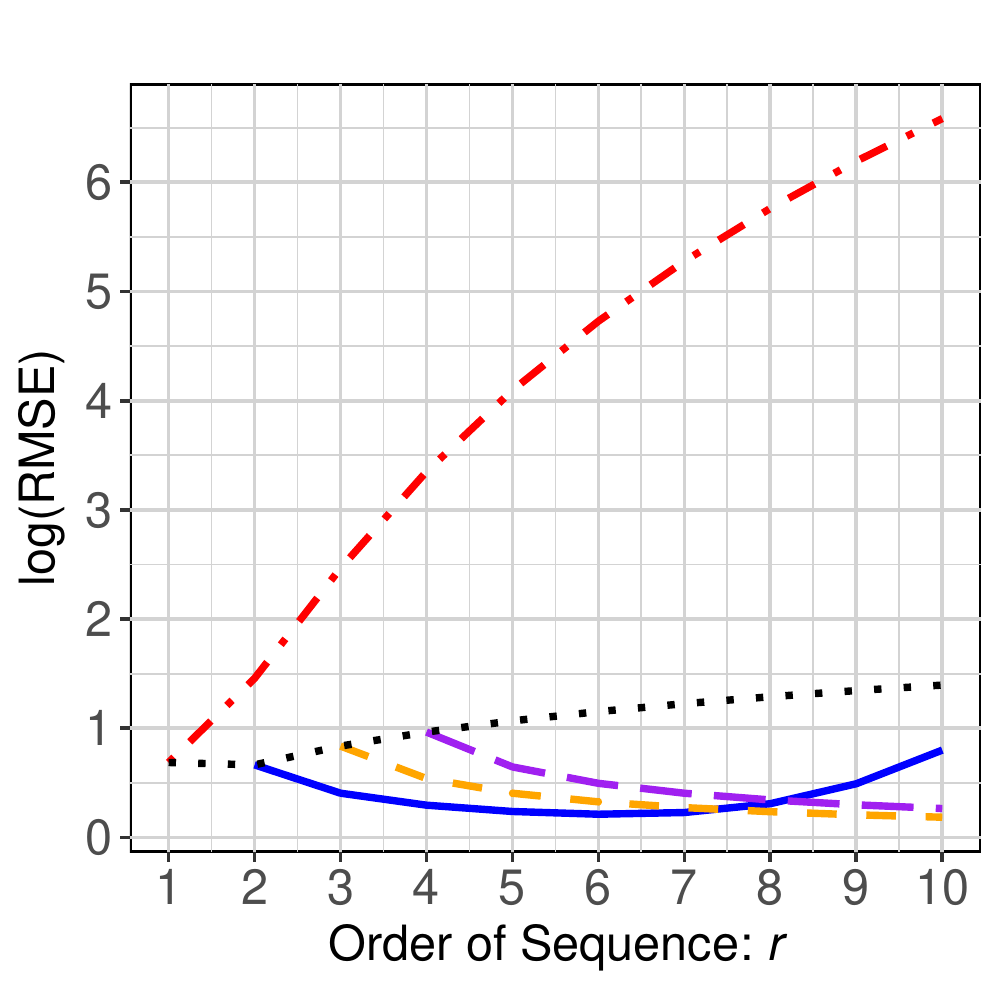} 
		\end{tabular}
		\caption{The logarithm of RVAR and RMSE for different sequences. The red dash-dotted curve: $d_{0}(r)$; the blue solid curve: $d_{1}(r)$; the yellow dashed curve: $d_{2}(r)$; the purple long-dashed curve: $d_{3}(r)$; the black dotted curve: $d_{r-1}(r)$. We consider the setting: $n=100$, $g(x)=5\sin(4\pi x)$, $x_i=i/n$ and $\varepsilon\sim N(0,1)$.}
		\label{delta-map}
	\end{figure}

	For further comparison, we also compute the numerical values of the asymptotic variance and bias according to Theorem \ref{optkr} for $\{(r,k): 1\le r \le 10,~0\le k \le {\rm min}(3,r-1)\}$, and then transform them into the relative variance (RVAR), squared bias (RSB), and the relative mean squared error (RMSE) by dividing the MSE by the scaling factor ${\rm var}(\varepsilon^2)/n$. As shown in Figure \ref{delta-map}, RVAR decreases to one for $\hat \sigma^2_{k}(r)$ with $0\le k \le 3$ but increases for $\hat \sigma^2_{r-1}(r)$ as $r$ becomes larger, which coincides with the functional pattern of $\delta_k(r)$. Due to the poorly-controlled estimation bias, $\hat \sigma^2_{0}(r)$ breaks down when $r$ is large. In contrast, thanks to a better balance on the bias-variance trade-off, the minimum RMSE for $\hat \sigma^2_{k}(r)$ with $1\le k \le 3$ is always smaller than that for the optimal and ordinary estimators.
	 To conclude, when estimating the residual variance, it is desirable to search for the whole optimal-$k$ family rather than restricting the attention only to the two extreme cases.

	\section{Simulation studies}

	\noindent
	This section compares the finite sample performance of the new and existing estimators within the optimal-$k$ family. 
	Noting that the optimal and ordinary estimators also belong to the optimal-$k$ family, to avoid confusion we refer to the optimal-$k$ estimators specifically as the optimal-$k$ estimators with $1\leq k\leq r-2$, i.e. the difference-based estimators associated with the blue points in Figure \ref{framework}. 
	By \citet{Dette1998}, an order of $r>4$ is rarely used in practice due to its complexity and also unstable performance. 
	This leads us to consider all the combinations of $(r,k)$ with $r\le4$, which yields 10 difference-based estimators in total, including the Rice estimator, three optimal estimators, three ordinary estimators, and three optimal-$k$ estimators.
	%for comparison, including the existing estimators and the new optimal-k estimators.
	%five existing estimators including $\hat \sigma^2_{\rm R}$, $\hat \sigma^2_{\text{opt}}(r)$ and $\hat \sigma^2_{\text{ord}}(r)$ with $r=2$ and $3$, and five new optimal-$k$ estimators with $(r,k)=(3,1),(4,1),(5,1),(4,2)~{\rm and}~(5,2)$.
	%In total, we have 10 difference-based estimators for the residual variance.

    To evaluate the performance of the difference-based estimators, we also follow the same mean function that is commonly used in the literature \citep{HallKayTit1990,Seifert1993,Dette1998}:
	\begin{align*}
		g(x) &=5{\sin}(w\pi x), \label{mean function}
		%g_2(x) &=5{\sin}(2\pi x),
		%g_2(x) &=5{\sin}(4\pi x).
		%g_3(x) &= 10\left[x+(2\pi)^{-1/2}\exp\left\{-100(x-0.5)^2\right\}\right].
	\end{align*}
	where $w>0$ controls the oscillation level of the mean function. Moreover, we let $x_i=i/n$ be equally spaced design points, $\{\varepsilon_i\}$ be a random sample of size $n$ from $N(0,\sigma^2)$, and $n=25$, $100$ and $500$ represent three different sample sizes. We further take $\{(\sigma,\omega): \sigma=0.2,0.4,\dots,2~{\rm and}~\omega=0.5,1\dots,5\}$, which forms a total of 100 combinations for the signal-to-noise ratios for the simulated data. 
	Finally, with 10000 simulations for each setting, we compute the RMSE for each estimator, defined as $(n/2\sigma^4){\rm MSE}$, and report the estimator with the smallest RMSE in Table \ref{Heatmap500}.

			\begin{table}[!t]
	\centering
		\caption{Optimal combination of $(r,k)$ under each of the 100 settings with the sample size $n=500$, 100 and 25. White cells: existing estimators; grey cells: new estimators. The $(r,k)$ value of the best option is provided in each cell.}
	\begin{tabular}{cc}
		\includegraphics[width=15cm]{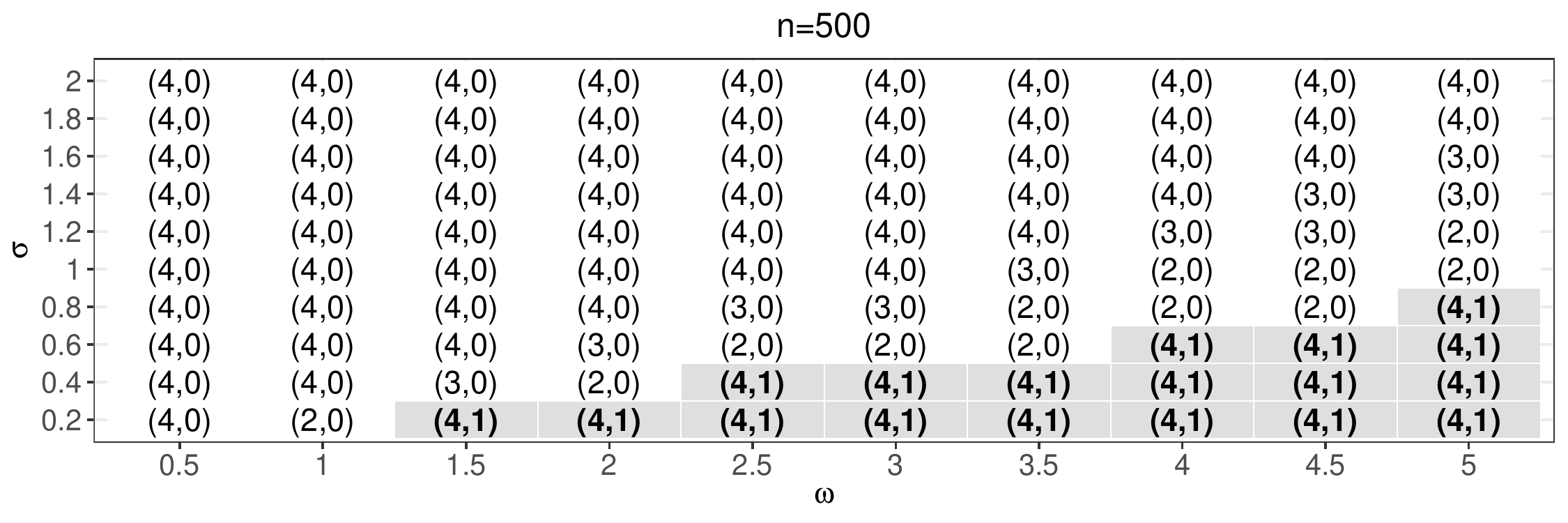} \\
		\includegraphics[width=15cm]{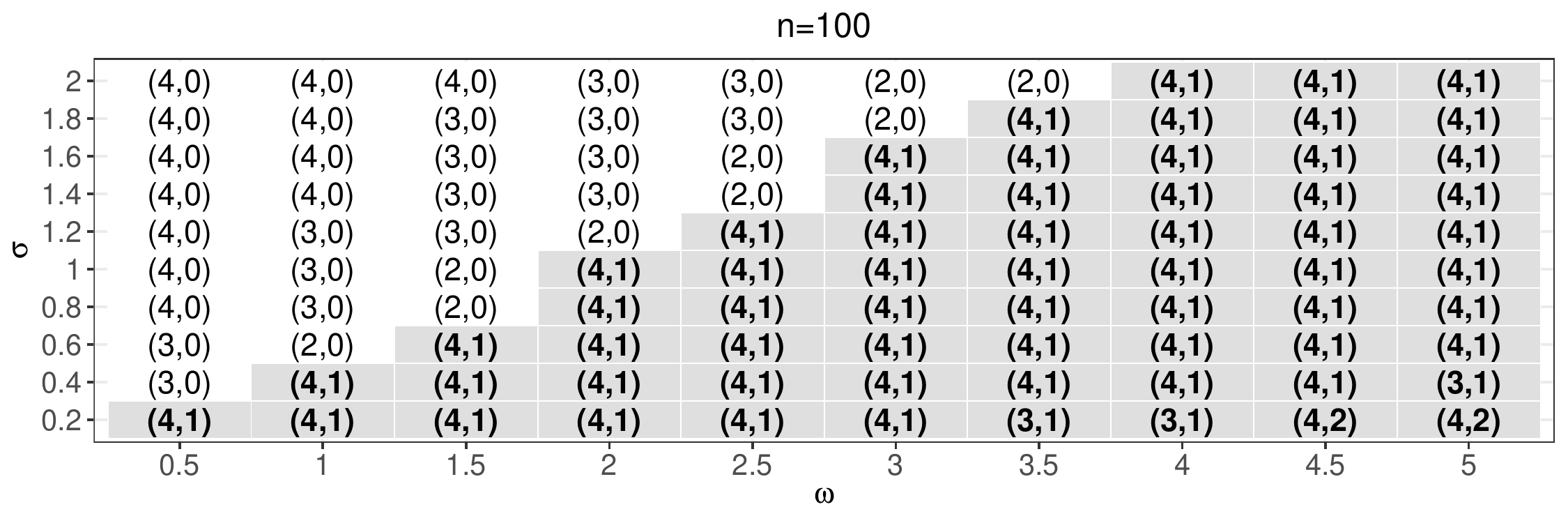}\\
		\includegraphics[width=15cm]{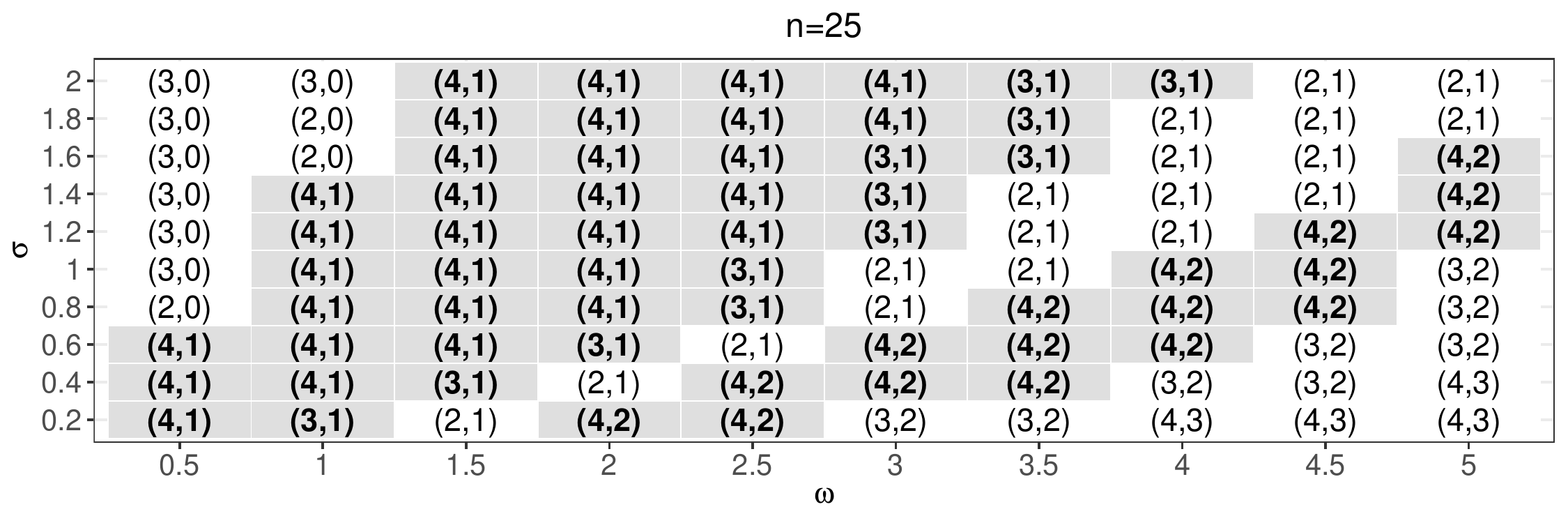}  
	\end{tabular}
	\label{Heatmap500}
\end{table}

From Table \ref{Heatmap500}, it is evident that the optimal-$k$ estimators provide the best performance in many settings, especially when the sample size is moderate to small. This coincides with the theoretical results, as well as the motivation, that the optimal-$k$ difference sequence offers a good balance between the estimation bias and the estimation variance. Moreover, the comparison results between the Rice,  optimal and ordinary estimators remain the same as observed in \citet{Dette1998}. Specifically, when the sample size is large, the ordinary estimators are often suboptimal due to the large estimation variance. In contrast, when the sample size is small, the optimal estimators tend to  be less satisfactory because of the uncontrolled estimation bias, especially when the mean function is also very rough. Lastly, the Rice estimator fails to provide the best performance in all of the settings. 
To sum up, with the newly introduced optimal-$k$ family, it has made possible for researchers to dramatically improve the existing difference-based estimation in nonparametric regression.

\begin{figure}[!t]
	\centering
	\begin{tabular}{cc}
		\includegraphics[width=15cm]{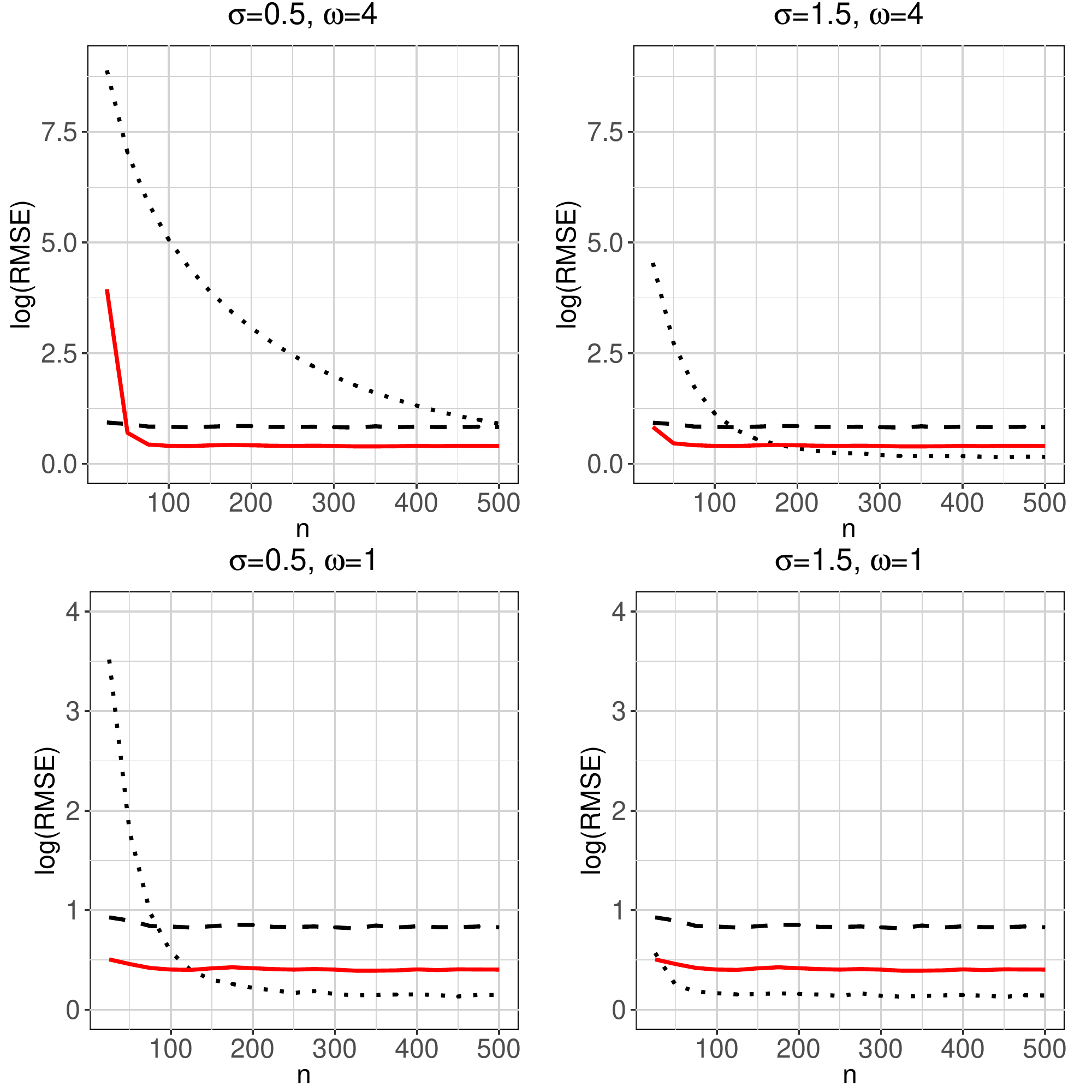} 
	\end{tabular}
	\caption{Plots of log(RMSE) as functions of $n$ for the three estimators with $r=3$: $\hat \sigma^2_{\text{opt}}(3)$ (dotted lines), $\hat \sigma^2_{\text{ord}}(3)$ (dashed lines) and $\hat \sigma^2_{1}(3)$ (red solid lines). The results are based on 10000 simulations.}
	\label{RMSE_vs_nn_r3}
\end{figure}

	\begin{figure}[!t]
	\centering
	\includegraphics[width=15cm]{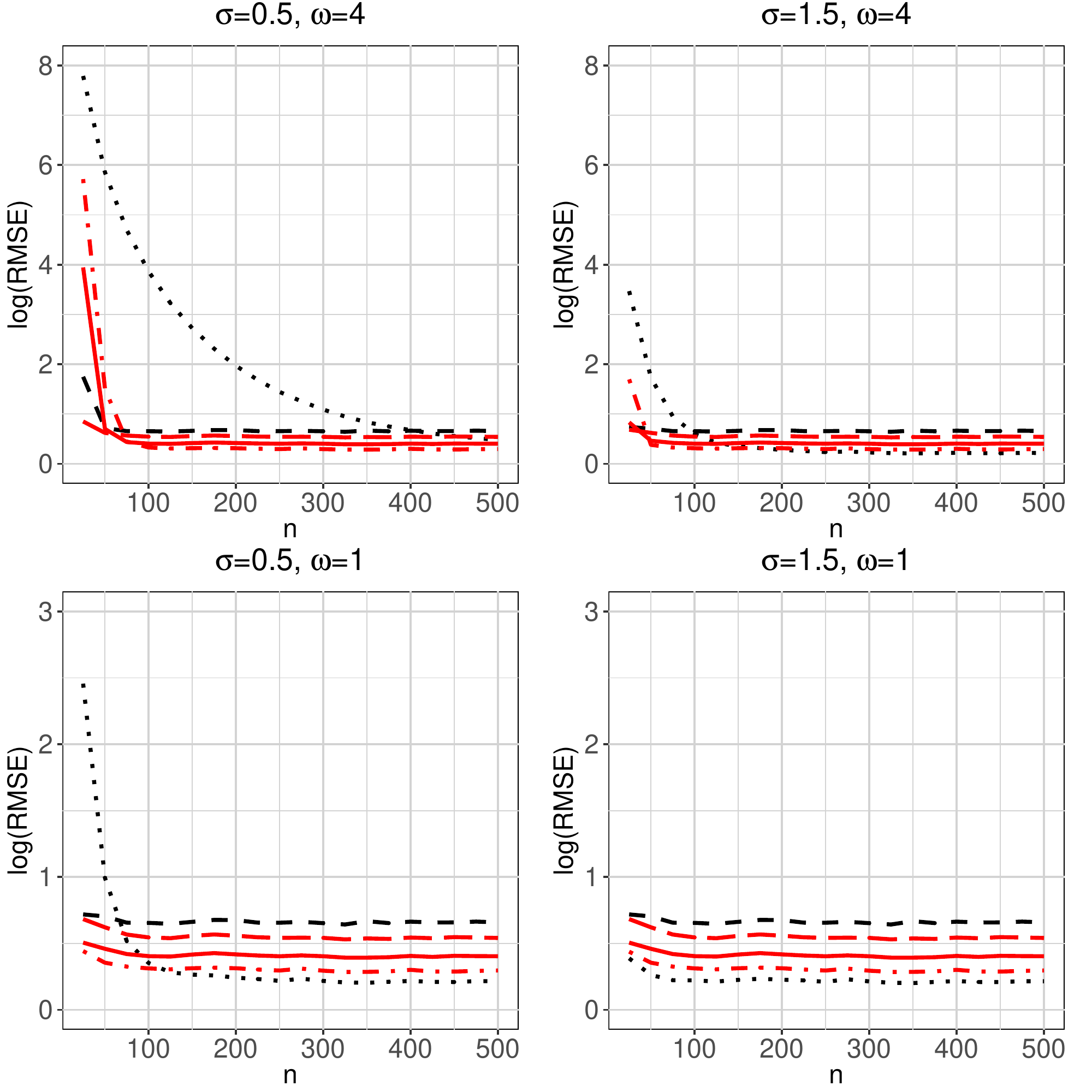}\\
	\caption{Plots of log(RMSE) as functions of $n$ for the five representative estimators with $2\le r \le 4$: $\hat \sigma^2_{\text{opt}}(2)$ (dotted lines), $\hat \sigma^2_{\text{ord}}(2)$ (dashed lines), $\hat \sigma^2_{1}(3)$ (red solid lines), $\hat \sigma^2_{1}(4)$ (red dash-dotted lines) and $\hat \sigma^2_{2}(4)$ (red long-dashed lines). The results are based on 10000 simulations.}
	\label{RMSE_vs_nn_r1234}
\end{figure}

To further explore the optimal-$k$ estimators, our second simulation is to conduct a cross-sectional study for assessing the effect of $k$ on the variance estimation when the order $r$ is given. 
We first consider $r=3$, which contains three difference-based estimators, namely the optimal estimator $\hat \sigma^2_{\text{opt}}(3)$, the ordinary estimator $\hat \sigma^2_{\text{ord}}(3)$, and the optimal-$k$ estimator $\hat \sigma^2_{1}(3)$. 
We further take $(\sigma, \omega)=(0.5,1)$, $(1.5,1)$, $(0.5,4)$ or $(1.5,4)$, and $n=25,50,\dots,500$. 
All other settings remain the same as before. As shown in Figure \ref{RMSE_vs_nn_r3} under the log scale, the optimal estimator is able to provide the smallest RMSE when the sample size is sufficiently large, but it tends to collapse easily when the sample size becomes smaller due to the poorly controlled estimation bias.
In contrast, the RMSE of the ordinary estimator is very stable along with the sample size. 
As a consequence, the ordinary estimator may not be optimal in the asymptotic sense, but it can provide a good estimate in the small sample size setting. 
Lastly, from Figure \ref{RMSE_vs_nn_r3}, it is also evident that the optimal-$k$ estimator performs consistently well in most settings and turns out to be the best choice among the three difference-based estimators with $r=3$. 

For additional information,  we also present the cross-sectional study with $r=4$ in Appendix B. From the results in Figure \ref{RMSE_vs_nn_r4}, it is evident again that both of the optimal-$k$ estimators outperform the two existing competitors. 
To further compare the two optimal-$k$ estimators, we note that $\hat\sigma_1^2(4)$ provides a smaller RMSE than $\hat\sigma_2^2(4)$ in most settings, showing that the bias correction with $k=1$ is usually enough for the difference-based estimation. 
But if, in practice, small sample sizes are a concern, then $\hat\sigma_2^2(4)$ can be preferred as well.

%reveals superiority only when the sample size is small and usually performs the worst when the sample size is moderate to large.The optimal-$k$ estimator maintains an outstanding balance between robustness and efficiency owing to its better bias-variance trade-offs. Based on the overall performance, we pick $\hat \sigma^2_{1}(3)$ as the best option of the order-3 estimators. Similar patterns also exist for the estimators of higher orders. As an example, we show the case of $r=4$ in Figure \ref{RMSE_vs_nn_r4} of Appendix B and recommend $\hat \sigma^2_{2}(4)$ for practical implementation for its robustness and high accuracy. We may conclude that the new optimal-$k$ estimators show superiority whenever order three or higher estimators are considered. 

Finally, to identify the best difference-based estimator with $r\leq 4$, we present the five local winners, including $\hat \sigma^2_{\text{opt}}(2)$, $\hat \sigma^2_{\text{ord}}(2)$, $\hat \sigma^2_{1}(3)$, $\hat \sigma^2_{1}(4)$ and $\hat \sigma^2_{2}(4)$, in Figure 4 under the log scale of RMSE. As expected, the three optimal-$k$ estimators all perform very well and they are also significantly better than the two estimators with $r=2$.  
%reveal similar performance and all outperform the two widely applied choices. Specifically, they are more robust than $\hat \sigma^2_{\text{opt}}(2)$, which is free of elimination for the estimation bias; meanwhile, they share similar robustness as $\hat \sigma^2_{\text{ord}}(2)$ but are more efficient in most cases. 
%that they achieve a better compromise between the estimation variance and the estimation bias.  
%Between the two new estimators, $\hat \sigma^2_{1}(3)$ is more efficient than $\hat \sigma^2_{2}(4)$ in most cases unless the sample size is small and the mean function is quite rough.
Among the three new estimators, we also note that the RMSE of $\hat \sigma^2_{1}(3)$ is always in the middle of those for the two optimal-$k$ estimators with $r=4$. In view of this, if we take into account both efficiency and robustness, $\hat \sigma^2_{1}(3)$ can be recommended as the final winner among all the candidate estimators. Another benefit for recommending $\hat \sigma^2_{1}(3)$ is that the second coefficient of  $d_1(3)=(0.2673,0,-0.8018,0.5345)$ is zero, and subsequently this estimator can be implemented as simply as the difference-based estimators with $r=2$. Taken together, we are now confident to recommend the optimal-$k$ difference sequence $d_1(3)$ for practical use, and for which it may also be considered as a rule of thumb.

	\section{Conclusion}
	\noindent
	Difference-based methods have been increasingly used in nonparametric regression, especially for estimating the residual variance. One fundamental problem, that is also of practical importance, is the difference sequence selection which has been studied for over three decades. The main contribution of this paper is to further advance the literature by introducing the optimal-$k$ difference sequence to better balance the bias-variance trade-off. 
	More importantly, it also dramatically enlarges the existing family of difference sequences that includes the optimal and ordinary difference sequences as two important special cases. 
	%Under the new framework, difference sequences of higher orders become plausible, which releases the restriction on the conventional options that the sequence order should not exceed four \citep{Dette1998}.
	To highlight a few key steps, we first reformulate the existing difference sequences as solutions to an optimization problem that minimizes the variance of the estimator under certain constraints. We then derive the optimal-$k$ difference sequence under the proposed optimization framework by providing more flexible constraints on the estimation bias and variance. The asymptotic properties of the optimal-$k$ estimator 
	are also established. Finally, we conduct extensive simulations to evaluate the difference-based estimators within the optimal-$k$ family, and more importantly identify the best difference sequence $d_1(3) = (0.2673, 0, -0.8018, 0.5345)$ for practical use.

Besides the new advances in finding the best difference sequence, another breakthrough in the difference-based methods is known to be the least squares estimator introduced by \citet{Tong2005} and \citet{Tong2013}. Specifically, they regressed the first order differences of paired observations on the squared distances between the paired covariates via a simple linear regression model, and then applied in spirit the simulation-extrapolation (SIMEX) method to estimate the error variance as the intercept. \cite{dai2017choice} extended their least squares estimator to form a unified framework by employing the higher-order differences among the observations as the regressors, and within the unified framework, they further concluded that the ordinary difference sequence can be consistently better than the optimal difference sequence. As an interesting future work, we will also incorporate the newly developed optimal-$k$ difference sequence to the unified framework and see whether, or to what extent, the new difference sequence will alter the conclusion made in \cite{dai2017choice}.

	Last but not least, to simplify the presentation of the main ideas, we have restricted our attention to the simple nonparametric regression model in (1) in the current paper. It is worth noting, however, that our proposed optimal-$k$ difference sequence is general and can also be applied to the difference-based estimation in many other models. To name a few such scenarios, the difference-based methods have been extended to estimate the residual variance in multivariate nonparametric regression models \citep{Munk2005}, in partial linear models \citep{wang2011difference}, and in nonparametric regression models with repeated measurements \citep{dai2015difference}. 			
	Other than the residual variance, the difference-based methods have also been effectively applied to estimate other quantities including, for example, the variance function or variogram \citep{bliznyuk2012variogram,yang2015variance},
	the autocovariance \citep{Hall2003,tecuapetla2016autocovariance,cui2021estimation}, 
	the error distribution \cite{chang2018a}, the derivatives of the mean function \citep{dai2016optimal,liu2018derivative,wang2019robust}, and the long-run variance in time series \citep{chan2022optimal}.
	Further research is needed to explore the potential applications of the optimal-$k$ difference sequence under these more general settings.
	%Another direction is to explore the performance of the optimal-$k$ sequence under the framework of least square variance estimator \cite{Tong2005,Tong2013,dai2017choice}.

	%\begingroup
	%%\setlength{\bibsep}{12pt}
	%\setstretch{1}
	%{
	%	\bibliographystyle{dcu}
	%	\citationstyle{dcu}
	%	\bibliography{reference_19Mar2014}
	%}
	%\endgroup
	%\bibliographystyle{Chicago}
	%\bibliographystyle{Chicago}
	%\bibliographystyle{asa}
	%\bibliographystyle{dcu}
	\bibliographystyle{apalike}

		%\citationstyle{dcu}
	\bibliography{optimal-k}

\begin{thebibliography}{}

\bibitem[Bliznyuk et~al., 2012]{bliznyuk2012variogram}
Bliznyuk, N., Carroll, R.~J., Genton, M.~G., and Wang, Y. (2012).
\newblock Variogram estimation in the presence of trend.
\newblock {\em Statistics and Its Interface}, 5:159--168.

\bibitem[Buckley et~al., 1988]{Buckley1988}
Buckley, M.~J., Eagleson, G.~K., and Silverman, B.~W. (1988).
\newblock The estimation of residual variance in nonparametric regression.
\newblock {\em Biometrika}, 75:189--199.

\bibitem[Carroll and Ruppert, 1988]{Carroll1988bk}
Carroll, R.~J. and Ruppert, D. (1988).
\newblock {\em Transformation and Weighting in Regression}.
\newblock London: Chapman and Hall.

\bibitem[Chan, 2022]{chan2022optimal}
Chan, K.~W. (2022).
\newblock Optimal difference-based variance estimators in time series: A
  general framework.
\newblock {\em The Annals of Statistics}, 50:1376--1400.

\bibitem[{Chang} et~al., 2018]{chang2018a}
{Chang}, J., {Delaigle}, A., {Hall}, P., and {Tang}, C.~Y. (2018).
\newblock A frequency domain analysis of the error distribution from noisy
  high-frequency data.
\newblock {\em Biometrika}, 105:353--369.

\bibitem[{Cui} et~al., 2021]{cui2021estimation}
{Cui}, Y., {Levine}, M., and {Zhou}, Z. (2021).
\newblock Estimation and inference of time-varying auto-covariance under
  complex trend: A difference-based approach.
\newblock {\em Electronic Journal of Statistics}, 15:4264--4294.

\bibitem[Dai et~al., 2015]{dai2015difference}
Dai, W., Ma, Y., Tong, T., and Zhu, L. (2015).
\newblock Difference-based variance estimation in nonparametric regression with
  repeated measurement data.
\newblock {\em Journal of Statistical Planning and Inference}, 163:1--20.

\bibitem[Dai et~al., 2016]{dai2016optimal}
Dai, W., Tong, T., and Genton, M.~G. (2016).
\newblock Optimal estimation of derivatives in nonparametric regression.
\newblock {\em Journal of Machine Learning Research}, 17(164):1--25.

\bibitem[Dai et~al., 2017]{dai2017choice}
Dai, W., Tong, T., and Zhu, L. (2017).
\newblock On the choice of difference sequence in a unified framework for
  variance estimation in nonparametric regression.
\newblock {\em Statistical Science}, 32:455--468.

\bibitem[Dette et~al., 1998]{Dette1998}
Dette, H., Munk, A., and Wagner, T. (1998).
\newblock Estimating the variance in nonparametric regression - what is a
  reasonable choice?
\newblock {\em Journal of the Royal Statistical Society, Series B},
  60:751--764.

\bibitem[Fan and Gijbels, 1996]{fan1996local}
Fan, J. and Gijbels, I. (1996).
\newblock {\em Local Polynomial Modelling and Its Applications}.
\newblock London: Chapman and Hall.

\bibitem[Gasser et~al., 1991]{Gasser1991}
Gasser, T., Kneip, A., and Kohler, W. (1991).
\newblock A flexible and fast method for automatic smoothing.
\newblock {\em Journal of the American Statistical Association}, 86:643--652.

\bibitem[Gasser et~al., 1986]{Gasser1986}
Gasser, T., Sroka, L., and Jennen-Steinmetz, C. (1986).
\newblock Residual variance and residual pattern in nonlinear regression.
\newblock {\em Biometrika}, 73:625--633.

\bibitem[Golub and Van~Loan, 1996]{golub1996matrix}
Golub, G.~H. and Van~Loan, C.~F. (1996).
\newblock {\em Matrix Computations, 3rd Ed}.
\newblock Baltimore: Johns Hopkins University Press.

\bibitem[Hall et~al., 1990]{HallKayTit1990}
Hall, P., Kay, J.~W., and Titterington, D.~M. (1990).
\newblock Asymptotically optimal difference-based estimation of variance in
  nonparametric regression.
\newblock {\em Biometrika}, 77:521--528.

\bibitem[Hall and Keilegom, 2003]{Hall2003}
Hall, P. and Keilegom, I.~V. (2003).
\newblock Using difference-based methods for inference in nonparametric
  regression with time series errors.
\newblock {\em Journal of the Royal Statistical Society, Series B},
  65:443--456.

\bibitem[Hall and Marron, 1990]{HallMarron1990}
Hall, P. and Marron, J.~S. (1990).
\newblock On variance estimation in nonparametric regression.
\newblock {\em Biometrika}, 77:415--419.

\bibitem[H{\"a}rdle, 1990]{hardle1990applied}
H{\"a}rdle, W. (1990).
\newblock {\em Applied Nonparametric Regression}.
\newblock Cambridge: Cambridge University Press.

\bibitem[Jiu and Li, 2021]{jiu2021hankel}
Jiu, L. and Li, Y. (2021).
\newblock Hankel determinants of certain sequences of {B}ernoulli polynomials:
  A direct proof of an inverse matrix entry from statistics.
\newblock {\em arXiv:2109.00772 \hskip -5pt}.

\bibitem[{Liu} and {Brabanter}, 2018]{liu2018derivative}
{Liu}, Y. and {Brabanter}, K.~D. (2018).
\newblock Derivative estimation in random design.
\newblock {\em Advances in Neural Information Processing Systems},
  31:3445--3454.

\bibitem[M{\"{u}}ller and Stadtm{\"{u}}ller, 1999]{Muller1999}
M{\"{u}}ller, H. and Stadtm{\"{u}}ller, U. (1999).
\newblock Discontinuous versus smooth regression.
\newblock {\em The Annals of Statistics}, 27:299--337.

\bibitem[Munk et~al., 2005]{Munk2005}
Munk, A., Bissantz, N., Wagner, T., and Freitag, G. (2005).
\newblock On difference-based variance estimation in nonparametric regression
  when the covariate is high dimensional.
\newblock {\em Journal of the Royal Statistical Society, Series B}, 67:19--41.

\bibitem[Rice, 1984]{Rice1984}
Rice, J. (1984).
\newblock Bandwidth choice for nonparametric regression.
\newblock {\em The Annals of Statistics}, 12:1215--1230.

\bibitem[Seifert et~al., 1993]{Seifert1993}
Seifert, B., Gasser, T., and Wolf, A. (1993).
\newblock Nonparametric estimation of residual variance revisited.
\newblock {\em Biometrika}, 80:373--383.

\bibitem[{Tecuapetla-Gómez} and {Munk}, 2016]{tecuapetla2016autocovariance}
{Tecuapetla-Gómez}, I. and {Munk}, A. (2016).
\newblock Autocovariance estimation in regression with a discontinuous signal
  and $m$-dependent errors: A difference‐based approach.
\newblock {\em Scandinavian Journal of Statistics}, 44:346--368.

\bibitem[Tong et~al., 2013]{Tong2013}
Tong, T., Ma, Y., and Wang, Y. (2013).
\newblock Optimal variance estimation without estimating the mean function.
\newblock {\em Bernoulli}, 19:1839--1854.

\bibitem[Tong and Wang, 2005]{Tong2005}
Tong, T. and Wang, Y. (2005).
\newblock Estimating residual variance in nonparametric regression using least
  squares.
\newblock {\em Biometrika}, 92:821--830.

\bibitem[Wand and Jones, 1995]{Wand1995bk}
Wand, M.~P. and Jones, M.~C. (1995).
\newblock {\em Kernel Smoothing}.
\newblock London: Chapman and Hall.

\bibitem[Wang et~al., 2011]{wang2011difference}
Wang, L., Brown, L.~D., and Cai, T.~T. (2011).
\newblock A difference based approach to the semiparametric partial linear
  model.
\newblock {\em Electronic Journal of Statistics}, 5:619--641.

\bibitem[{Wang} et~al., 2019]{wang2019robust}
{Wang}, W.~W., {Yu}, P., {Lin}, L., and {Tong}, T. (2019).
\newblock Robust estimation of derivatives using locally weighted least
  absolute deviation regression.
\newblock {\em Journal of Machine Learning Research}, 20(60):1--49.

\bibitem[Wang, 2011]{Wang2011bk}
Wang, Y. (2011).
\newblock {\em Smoothing Splines: Methods and Applications}.
\newblock New York: Chapman and Hall.

\bibitem[Yang and Zhu, 2015]{yang2015variance}
Yang, S. and Zhu, Z. (2015).
\newblock Variance estimation and kriging prediction for a class of
  non-stationary spatial models.
\newblock {\em Statistica Sinica}, 25:135--149.

\bibitem[Yatchew, 2003]{yatchew2003semiparametric}
Yatchew, A. (2003).
\newblock {\em Semiparametric Regression for the Applied Econometrician}.
\newblock New York: Cambridge University Press.

\end{thebibliography}

	\newpage
	\section*{Appendix A}

	%\vskip 15pt
	%\noindent
	%\textbf{Proof of Lemma 1.}

	\vskip 15pt
	\noindent
	%{\bf Proof of Theorem \ref{delta-kr}.}\\
	To prove the main results, we first introduce a lemma on the equivalence of two groups of constraints, by which we can transform the nonlinear optimization problem (\ref{opt-k}) into a linear optimization problem so that a closed-form solution for minimizing $\delta(r)$ can be derived. 
	
	%To prove the main results, we first introduce two lemmas. The first lemma is on the equivalence of two groups of constraints, by which we can transform the nonlinear optimization problem (\ref{opt-k}) into a linear optimization problem so that a closed-form solution for minimizing $\delta(r)$ can be derived. The second lemma shows that a matrix involved in Theorem~1 is non-singular. %The third lemma shows the equivalence between $d_{r-1}(r)$ and the ordinary difference sequence.
	
	\vskip 15pt
	\begin{lemma} Assume that $C_0=0$. Then for any $1\le k \le r-1$,  the constraints $C_1=\dots=C_k=0$ are equivalent to 
		$$\sum_{c=1}^r c^{2s}D_c=0,\quad  s=1,\dots, k,$$
		where $D_c=\sum_{j=0}^{r-c}d_jd_{j+c}$, for $c=1,\dots,r$, are the same as defined in Theorem 2.
	\end{lemma}
	
	Proof of Lemma 1. By the definition of $D_c$, for any integer $s\ge 1$, we have
	\begin{eqnarray}
		\sum_{c=1}^r c^{2s}D_c&=&\sum_{c=1}^r \sum_{j=0}^{r-l}c^{2s}d_jd_{j+c}
		=\sum_{k_1-k_2=1}^r (k_1-k_2)^{2s}d_{k_1}d_{k_2}\nonumber \\
		&=&\sum_{k_1-k_2=1}^r \left[\sum_{t=0}^{2s}{2s \choose t}k_1^{t}k_2^{2s-t}(-1)^{2s-t}\right]d_{k_1}d_{k_2}\nonumber \\
		&=&\sum_{t=0}^{2s}(-1)^{2s-t}{2s \choose t}\sum_{k_1-k_2=1}^r k_1^{t}k_2^{2s-t}d_{k_1}d_{k_2}\nonumber \\
		&=&\frac{1}{2}\sum_{t=0}^{2s}(-1)^{2s-t}{2s \choose t}\left(\sum_{k_1=0}^r k_1d_{k_1}^{t}\right)\left(\sum_{k_2=0}^r k_2d_{k_2}^{2s-t}\right)\nonumber \\
		&=&\frac{1}{2}\sum_{t=0}^{2s}(-1)^{2s-t}{2s \choose t}C_tC_{2s-t} t!(2s-t)!. \label{AP4}
	\end{eqnarray}	
	
	\vskip 10pt
	%\noindent
	%From ``$C_0=\dots=C_k=0$'' to ``$\sum_{c=1}^r c^{2s}D_c=0,~{\rm for} ~s=1,\dots,k.$"
	\vskip 10pt
	To show the equivalence, we first assume that $C_1=\dots=C_k=0$ holds. Then for any $s\le k$, we have ${\rm min}(t,2s-t)\le s\leq k$ for any given $t$. Consequently, 
	$$C_t=0 ~\quad {\rm or} \quad ~C_{2s-t}=0,\quad t=1,\dots,2s-1.$$
	Moreover, by (\ref{AP4}) and the assumption $C_0=0$, it yields that $\sum_{c=1}^r c^{2s}D_c=0$ for $s=1,\dots, k.$
	\vskip 10pt
	We now assume that $\sum_{c=1}^r c^{2s}D_c=0$ holds for $s=1,\dots, k$. In what follows, we show that $C_1=\dots=C_k=0$ by induction.
	\begin{enumerate}
		\item [a)] When $k=1$, we have $\sum_{c=1}^r c^{2}D_c=0$. Then by (\ref{AP4}), it yields that
		$$0=\sum_{c=1}^r c^{2}D_c=-\left(\sum_{k_1=0}^r k_1d_{k_1}\right)\left(\sum_{k_2=0}^r k_2d_{k_2}\right)=-\left(\sum_{k=0}^r k_1d_{k_1}\right)^2,$$
		which leads to $\sum_{k=0}^r k_1d_{k_1}=0$, i.e. $C_1=0$. 
		\vskip 5pt
		\item [b)] Assume that the conclusion holds for $k=\tilde k$. That is, when $\sum_{c=1}^r c^{2s}D_c=0~{\rm for}~s=1,\dots,\tilde k$, it follows that $C_1=\dots=C_{\tilde k}=0$. Then for $k=\tilde k+1$, by (\ref{AP4}) and $\sum_{c=1}^r c^{2({\tilde k}+1)}D_c=0$, we have
		\begin{equation}
			\frac{1}{2}\sum_{t=0}^{2(\tilde k+1)}(-1)^{2(\tilde k+1)-t}{2(\tilde k+1) \choose t}C_tC_{2(\tilde k+1)-t}t!(2(\tilde k+1)-t)!=0. \label{L2_1}
		\end{equation}
		Moreover, by the assumptions that $C_0=C_1=\dots=C_{\tilde k}=0$, (\ref{L2_1}) can be simplified as 
		\begin{equation*}
			\sum_{c=1}^r c^{2({\tilde k}+1)}D_c=\frac{1}{2}(-1)^{\tilde k+1}{2(\tilde k+1) \choose \tilde k+1}C^2_{\tilde k+1}((\tilde k+1)!)^2=0. 
		\end{equation*}
		This shows that $C_{\tilde k +1}=0$ and thus the conclusion also holds for $k=\tilde k+1$. This completes the proof of Lemma 1.  
		%    show that when $k=\tilde k+1$, $C_0=\dots=C_{\tilde k}=0$ and $\sum_{l=1}^r c^{2s}D_c=0,~{\rm for}~s=1,\dots, {\tilde k}+1$. 
	\end{enumerate}

	{\centering {\em Proof of Theorem 1}
		
	}
	(a) To prove the invertibility of $V_k$, we first consider $k=r-1$ and decompose the matrix as $V_{r-1}={U}_r^{T}{U}_r$, where $U_r$ is the $r\times r$ Vandermonde matrix with form
	\begin{equation*}
		{U}_r=\left(
		\begin{array}{ccccc}
			1~ & ~1    &~ 1  &~\cdots &~ 1\\
			1~ &~ 2^2  &~ 2^4&~\cdots &~ 2^{2(r-1)}\\
			1~ & ~3^2  & ~3^4&~\cdots &~ 3^{2(r-1)}\\
			\vdots~ &~\vdots&~ \vdots&~\ddots &~\vdots \\
			1~ & ~r^2  &~ r^4&~\cdots&~ r^{2(r-1)}\\
		\end{array}
		\right).
	\end{equation*}
	The decomposition holds by noting that 
	$$V_{r-1}(i,j)=I_{2(i+j)-4}=\sum_{l=1}^r l^{2(i-1)}l^{2(j-1)}={U}_r(\cdot,i)^{T}{U}_r(\cdot,j),$$
	where ${U}_r(\cdot,i)$ denotes the $i$th colomn of $U_r$. 
	According to the properties of Vandermonde matrix \citep{golub1996matrix,jiu2021hankel}, we further have 
	$$|{U}_r|=\prod_{1\le i<j\le r} (j^2-i^2) \neq 0,$$
	so that ${U}_r$ is an invertible matrix. This shows that $V_{r-1}$ is a positive-definite matrix with all the principal submatrices invertible; that is $|V_k|\neq 0$ for any $0\le k\le r-1$. %This completes the proof of Lemma 2.
	
	\vskip 10pt
	(b) For the equidistant design, when $k\ge 1$ in (\ref{opt-k}), the optimization problem is given as
	\begin{eqnarray}
		\underset{{d(r)\in \mathbb{R}^{r+1}}}{\rm arg~min}~\delta(r) \quad {\rm subject~to}\quad C_0=\dots=C_k=0~ {\rm and}~\sum_{j=0}^r d_j^2=1. \label{optimal-k}
	\end{eqnarray}
	By Lemma 1, the constraints in (\ref{optimal-k}) can be equivalently expressed as
	$$\sum_{c=1}^r c^{2s}D_c=0,~s=1,\dots,k, \quad{\rm and}\quad \sum_{c=1}^r D_c=-\frac{1}{2}\sum_{j=0}^r d_j^2=-\frac{1}{2}.$$
	%{\color{blue}{Sorry, I fogot to interpret this in the previous version. Constraints in (\ref{optimal-k}) consist of two parts: (I) $C_1=\dots=C_k=0$, (II) $C_0=0$ and $\sum_{j=0}^r d_j^2=1$. (I) is equivalent to $\sum_{c=1}^r c^{2s}D_c=0,~s=1,\dots,k$, and (II) is equivalent to $\sum_{c=1}^r D_c=-\frac{1}{2}\sum_{j=0}^r d_j^2$ (taking squares of $C_0=0$) and $\sum_{j=0}^r d_j^2=1$, which lead to $\sum_{c=1}^r D_c=-\frac{1}{2}$. So, I think it is fine to present in the current form.}}
	
	Consequently, the optimization problem becomes
	\begin{eqnarray*}
		\underset{{d(r)\in \mathbb{R}^{r+1}}}{\rm arg~min}~\sum_{c=1}^r D_c^2\quad {\rm subject~to}\quad \sum_{c=1}^r c^{2s}D_c=0,~s=1,\dots,k,~{\rm and}~\sum_{c=1}^r D_c=-\frac{1}{2}.
	\end{eqnarray*}
	
	To apply the method of Lagrange multipliers, we let
	\begin{eqnarray*}
		L(D_1,\dots,D_r,\lambda_0,\dots,\lambda_k)=\sum_{c=1}^r D_c^2+\lambda_0(\sum_{c=1}^r D_c+\frac{1}{2})+\sum_{s=1}^k \left(\lambda_{s}\sum_{c=1}^r c^{2s}D_c\right).
	\end{eqnarray*}
	Then by taking the partial derivatives of $L$ and setting them as zero, we have
	\begin{eqnarray}
		{\frac{\partial L}{\partial D_c}}&=&2D_c+\lambda_0+\sum_{s=1}^k c^{2s}\lambda_{s}=0, \quad c=1,\dots,r,\label{opt_con1}\\
		\frac{\partial L}{\partial \lambda_0}&=&\sum_{c=1}^r D_c+\frac{1}{2}=0, \label{opt_con2}\\
		\frac{\partial L}{\partial \lambda_{s}}&=&\sum_{c=1}^r c^{2s}D_c=0, \quad s=1,\dots,k.\label{opt_con3}
	\end{eqnarray}
	Moreover, by taking the weighted sum of the $r$ equations in (19), it yields that 
	\begin{eqnarray*}
		\sum_{c=1}^r c^{2t}\frac{\partial L}{\partial D_c}=2\sum_{c=1}^r c^{2t}D_c+\lambda_0\sum_{c=1}^r c^{2t}+\sum_{s=1}^k\lambda_{s} \sum_{c=1}^r c^{2(s+t)}=0, \quad t=1,\dots,k.
	\end{eqnarray*}
	By (\ref{opt_con2}) and (\ref{opt_con3}), the above equations can be expressed as
	\begin{eqnarray}
		{V}_{k}{\Lambda}={e}_1, \label{mateq}
	\end{eqnarray}
	where
	${\Lambda}=(\lambda_0,\lambda_1,\dots,\lambda_k)^{T}$ and $e_1=(1,0,\dots,0)^T.$
	By solving Equation (\ref{mateq}), we get
	\begin{eqnarray}
		{\Lambda}=({V}_{k}^{-1}(1,1),\dots,{V}_{k}^{-1}(k+1,1))^{T}. \label{lambda}
	\end{eqnarray}
	%and then it is straightforward to calculate $D_c$'s according to (\ref{opt_con1}).
	Further by combining (\ref{opt_con1}), (\ref{opt_con2}) and (\ref{opt_con3}), we get the minimum value of $\delta(r)$ as
	\begin{eqnarray*}
		\delta_k(r)&=&\sum_{c=1}^r D_c^2=-\frac{1}{2}\sum_{c=1}^r D_c(\lambda_0+\sum_{s=1}^k c^{2s}\lambda_{s})\\
		&=&-\frac{1}{2}\lambda_0\sum_{c=1}^r D_c-\frac{1}{2}\sum_{s=1}^k\lambda_{s} \sum_{c=1}^r c^{2s}D_c\\
		&=&\frac{\lambda_0}{4}.
	\end{eqnarray*}
	By (\ref{lambda}), we have $\lambda_0={V}_{k}^{-1}(1,1)$ and hence $\delta_k(r)={V}_{k}^{-1}(1,1)/4$.
	
	\vskip 15pt

	{\centering 	{\em Proof of Theorem \ref{calculation}}
		
	}
	
	To generate the optimal-$k$ difference sequence, we first construct a polynomial $R(t)$ in the following form
	\begin{eqnarray*}
		R(t)&=&t^r\{D_r(t^r+t^{-r})+\cdots+D_1(t+t^{-1})+1\}\\
		&=&	t^r(d_0+d_1t^1+\dots+d_rt^r)(d_0+d_1t^{-1}+\dots+d_rt^{-r}).
	\end{eqnarray*}
	$R(t)$ is a self-reciprocal polynomial, which means that if $z$ is a root of $R(t)$ then $1/z$ must also be a root of $R(t)$. Also, if a complex number $z$ is a root of $R(t)$ then its conjugate $\bar z$ is also a root of $R(t)$ since $D_c$'s are real.

	In $R(t)$, the coefficients $D_c$'s can be calculated according to Theorem~\ref{delta-kr} and hence we can explicitly get all the $2r$ roots of $R(t)$ via any root-finding algorithm.
	Note that $\sum_{c=1}^{r}D_c=-1/2$, so we know that $t=1$ is a double root of $R(t)$ and the $2r$ root of $R(t)$ can be expressed as  $$RT=\{z_1,z_2,\dots,z_r,z_1^{-1},z_2^{-1},\dots,z_r^{-1}\}~ {\rm with}~ z_1=z_1^{-1}=1.$$
	
	To restore the polynomial $d_0+d_1t+\dots+d_rt^r$ and hence get the desired difference sequence as the normalized coefficients, we adopt the following two criteria to select $r$ roots from $RT$:
	(i) Choose any set of $r$ roots with no repeat of the index. (ii) If a complex root $z\in RT$ is selected, then its conjugate $\bar z\in RT$ should also be selected.
	The first criterion is to ensure that the restored polynomial is proportional to the desired one, i.e.,
	$$\prod_{i=1}^{r}(t-z_i)=a_0+a_1t+\dots+a_rt^r={(a_0^2+\cdots+a_r^2)^{1/2}}(d_0+d_1t+\dots+d_rt^r).$$
	The second criterion is to ensure that the coefficients of the restored polynomial are real numbers.
	
	Various valid sets of roots exist, among which we recommend simply separating the roots with the unit circle and choosing the roots on or outside the circle as described in Theorem \ref{calculation}, which satisfies the two criteria. 
	\vskip 15pt
	
	{\centering {\em Proof of Theorem \ref{optkr}}
		
	}
	
	The asymptotic bias of $\hat \sigma_k^2(r)$ is an immediate result from equation (\ref{bias-d}).
	While for the asymptotic variance, according to (\ref{quadratic mse}), we have 
	\begin{align}\label{T3_var}
		{\rm var}(\hat \sigma_k^2(r))=&~{1\over (n-r)^2}\left[4\sigma^2g^{T}D^2g+4g^{T}\{D{\rm diag}(D)u\}\sigma^3\gamma_3\right.\nonumber \\
		&\left.+~\sigma^4{\rm tr}\{{\rm diag}(D)^2\}(\gamma_4-3)+2\sigma^4{\rm tr}(D^2)\right].
	\end{align}
	By \cite{Dette1998}, together with the definition of $D$ and the assumption that the mean function has a continuous $r$th derivative,we can derive that
	\begin{eqnarray*}
		g^{T}D^2g&=&\|Dg\|_2^2=O(1),\\
		g^{T}\{D{\rm diag}(D)u\}&=&O(1),\\
		{\rm tr}\{{\rm diag}(D)^2\}&=&n+o(n),\\
		{\rm tr}(D^2)&=&n\{1+2\delta(r)+o(1)\}.
	\end{eqnarray*}
	Plugging them back to (\ref{T3_var}), 
	$${\rm var}(\hat \sigma_k^2(r))={\frac{1}{n}}\{{\rm var}(\varepsilon^2)+4\sigma^4\delta(r)\}+o(\frac{1}{n}).$$
	This proves the theorem by noting that $\delta(r)={V}_{k}^{-1}(1,1)/4$ from Theorem~1.
	
	\section*{Appendix B}

	\vskip 10pt
	\noindent
This appendix presents the cross-sectional study with $r=4$, which includes a total of 4 estimators including the optimal estimator $\hat \sigma^2_{\text{opt}}(4)$, the ordinary estimator $\hat \sigma^2_{\text{ord}}(4)$, and two optimal-$k$ estimators $\hat \sigma^2_{1}(4)$ and $\hat \sigma^2_{2}(4)$. 
For a fair comparison, we follow the same settings as in Figure \ref{RMSE_vs_nn_r3} and report the logarithm of RMSE for each estimator in Figure \ref{RMSE_vs_nn_r4}. 

%Both the optimal and ordinary estimators perform similarly as the case when $r=3$; see Figure \ref{RMSE_vs_nn_r3}. The two optimal-$k$ estimators are still superior owing to their robust and accurate estimation across most settings. Compared with $\hat \sigma^2_{1}(4)$, $\hat \sigma^2_{2}(4)$ is more robust when the sample size is small and reaches similar accuracy when the sample size is moderate to large. Hence, we may pick $\hat \sigma^2_{2}(4)$ as the representative of the estimators with $r=4$. 
\setcounter{figure}{0}
\renewcommand{\thefigure}{S\arabic{figure}}

	\begin{figure}[!b]
	\centering
	\begin{tabular}{cc}
		\includegraphics[width=15cm]{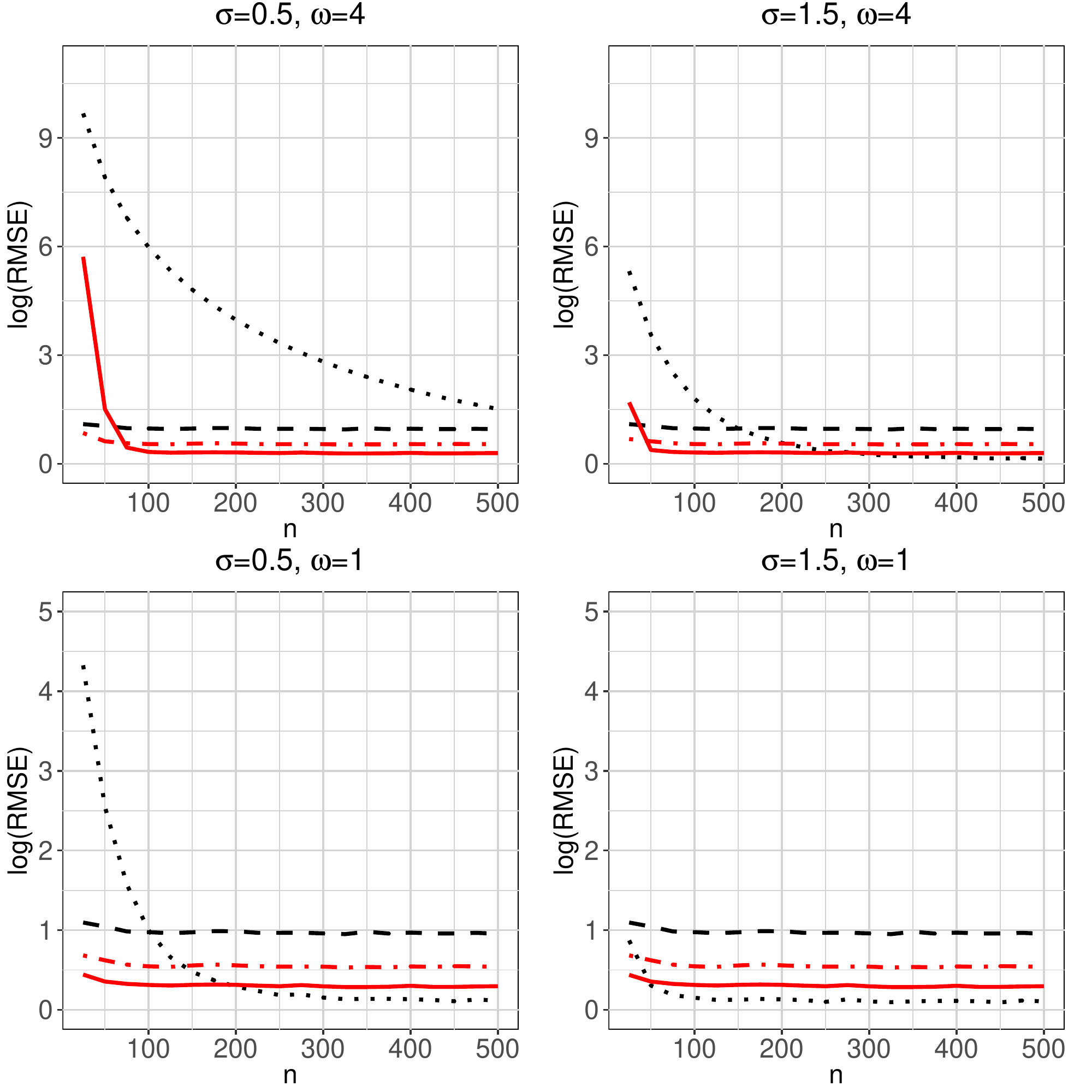} 
	\end{tabular}
	\caption{Plots of log(RMSE) as functions of $n$ for the four estimators with $r=4$: $\hat \sigma^2_{\text{opt}}(4)$ (dotted lines), $\hat \sigma^2_{\text{ord}}(4)$ (dashed lines), $\hat \sigma^2_{1}(4)$ (red solid lines) and $\hat \sigma^2_{2}(4)$ (red dash-dotted lines). The results are based on 10000 simulations.}
	\label{RMSE_vs_nn_r4}
\end{figure}

\end{document}